\pdfoutput=1
\documentclass{article}
\usepackage{amsmath}
\usepackage{amsfonts}
\usepackage{bm}
\usepackage{graphicx}
\usepackage[a4paper, total={6in, 8in}]{geometry}
\usepackage[colorinlistoftodos]{todonotes}
\usepackage{hyperref}
\hypersetup{colorlinks=true, allcolors=blue}
\usepackage{float}
\usepackage{cases}
\usepackage{amsmath}
\usepackage{amssymb}
\usepackage{amsbsy}
\usepackage{bm}
\usepackage{siunitx}
\usepackage[toc,page]{appendix}
\usepackage{psfrag}
\usepackage{pdfpages}
\usepackage{mathrsfs}
\usepackage{wrapfig}
\numberwithin{equation}{section}
\DeclareMathOperator{\tr}{tr}
\newcommand*\diff{\mathop{}\!\mathrm{d}}
\newcommand{\me}{\mathrm{e}}
\newcommand{\mi}{\mathrm{i}}
\definecolor{ao(english)}{rgb}{0.0, 0.5, 0.0}
\usepackage{caption}
\usepackage{subcaption}
\usepackage{upgreek}
\usepackage[utf8]{inputenc}
\newcommand\numberthis{\addtocounter{equation}{1}\tag{\theequation}}

\begin{document}



\title{Stress relaxation and thermo-visco-elastic effects in fluid-filled slits and fluid-loaded plates}

\author{Erik~Garc\'ia~Neefjes${}^{1}$, David~Nigro${}^{2}$,
Rapha{\"e}l~C.~Assier${}^{3}$,\\ William~J.~Parnell${}^{3}$
\\[5pt]
{\footnotesize
${}^{1}$ School of Mathematical and Physical Sciences, Macquarie University, Sydney NSW 2109, Australia}
\\
{\footnotesize
${}^{2}$Thales United Kingdom, 350 Longwater Avenue Green Park, Reading RG2 6GF, UK}
\\
{\footnotesize
${}^{3}$Department of Mathematics, University of Manchester, Oxford Rd, Manchester, M13 9PL, UK}
}

\date{\today}

\maketitle
\begin{abstract}
In this paper, we theoretically analyse wave propagation in two canonical problems of interest: fluid-filled thermo-visco-elastic slits and fluid-loaded thermo-visco-elastic plates. We show that these two configurations can be studied {via} the same pair of dispersion equations with the aid of the framework developed in \cite{garcia2022unified}, which incoporates thermal effects. These two problems are further interrelated, since in the short wavelength limit (relative to the slit/plate width) the respective modes are governed by the same dispersion equation, commonly known as the Scholte--Stoneley equation. It is the Scholte-type modes {that} are mainly analyzed in this paper. {We illustrate results when the fluid is water, although the theory is valid for any Newtonian fluid.} 
Both `hard' and `soft' solids are compared, with the emphasis being  placed on the importance of thermo-viscoelastic effects, particularly when \textit{stress relaxation} is considered. Two main recent works are discussed extensively, namely \cite{cotterill2018thermo} for slits and \cite{staples2021coupled} for loaded plates, both of which do not incorporate viscoelastic mechanisms. We show how the consideration of viscoelasticity can extend the results discussed therein, and explain the circumstances under which they arise. 

\end{abstract}

\section{Introduction}

{Acoustic and elastic wave propagation in confined ducts, channels/slits and in plates is a classical field with a broad range of applications \cite{rose2014guided, rienstra2015fundamentals}. Loss mechanisms in a variety of such geometries are still of significant interest from a research perspective, particularly when these configurations are employed as the microstructures in metamaterials. As an example, partly motivated by experimental work in \cite{ward2015boundary} for air-filled channels, the work in \cite{cotterill2018thermo} considered acoustic propagation in water-filled steel slits. Specifically the study focused on the importance of boundary layer attenuation in the fluid region as the slits become narrow.} It was found that although the presence of fluid viscosity is necessary to describe the attenuation of the mode along the slit, in order to capture the large reduction in phase speed as the width decreases, it is key to capture the fluid-structure interaction (FSI) effects, as is well known in underwater acoustics. On the other hand, for air-filled slits, FSI may be ignored in most instances, even for considerably soft media such as rubbers \cite{garcia2022unified}, but instead, thermal dissipative effects should be taken into account \cite{kirchhoff1868ueber, bruneau2013fundamentals} as opposed to water under normal  circumstances \cite{qi1994attenuated}.

{One of the main objectives of this paper is to extend the analysis in \cite{cotterill2018thermo} to the case of \textit{soft} solid slits/plates and in particular to take into account viscoelastic losses in such soft solids.} Following the discussion above, in order to exploit these mechanisms for fluid-filled slits, it is required to {have strong FSI effects} and therefore only results for water are considered, although the theory is applicable to any Newtonian fluid. The fluid-solid half-space which has been widely studied under many circumstances \cite{gusev2015effect, zhu2004leaky, favretto1996theoretical, glorieux2001character}, is considered first since it constitutes the geometrical limit as the channel width increases (relative to the wavelength of the transverse wave in the fluid). This configuration gives rise to two well known main families of interface wave solutions corresponding to \textit{Leaky--Rayleigh} (LR) and \textit{Stoneley--Scholte} (Sc) modes. Unlike the LR \cite{mozhaev2002subsonic}, the Sc mode propagates for any fluid-solid interface and in particular is the only propagating mode for water-soft interfaces \cite{glorieux2001character}, which we consider here. The Sc phase speed is reduced significantly for `soft' fluid-solid interfaces \cite{staples2021coupled} as opposed to `hard' fluid-solid interfaces where the Sc phase speed is approximately equal to the speed of sound in the fluid.
The influence of viscoelasticity on this mode for water-soft media (synthetic resins) is studied theoretically and experimentally in a series of papers \cite{favretto1996theoretical,favretto1997excitation,favretto1999identification} with the objective of using Sc waves to characterise the properties of the ocean's sedimentary bottom. 

In the current work, we bring special attention to the importance of viscoelastic \textit{stress relaxation} in the phase-speed and attenuation of the Sc mode (and body waves) and the corresponding regimes where these effects become important. Under the standard linear solid model (SLSM) used here, the key dimensionless parameter in the frequency domain is the \textit{Deborah number} $\omega t_r$ where $\omega$ is the angular frequency and $t_r$ the single relaxation time of the material under consideration \cite{garcia2022unified}. Although the relaxation times of soft materials can vary over several orders of magnitude \cite{zajac2017relaxation}, the frequencies involved in the majority of reported studies are such that $\omega t_r \gg 1 $, which we presume is the reason for the lack of stress relaxation analyses in the form presented in this paper. Despite this, we believe the regime considered here remains of strong interest since these effects are clearly visible in experiments \cite{liao2006estimation}, and the importance of stress relaxation is often reported in the literature \cite{staples2021coupled, hui2016effect}.\\

Another important part of this paper is devoted to the study of (thermo-viscous) fluid-loaded plates. {In spite of the fact that this configuration is physically different to the case of fluid-filled slits, the associated dispersion equations (DEs) are in fact identical in both problems thanks to the generality of the media considered \cite{garcia2022unified}.} Early work on fluid-loaded plates \cite{osborne1945transmission,schoch1952schalldurchgang} identified that the presence of the liquid causes the standard plate Lamb wave solutions to become \textit{leaky} (as for Rayleigh modes in the half-space) and another two solutions arise which are not present in the absence of the fluid. These solutions are in fact Scholte-like interface waves which become coupled in the plate region as the thicknesses decreases, and in order to distinguish them from their half-space counterpart we define them as `coupled plate--Scholte' modes \cite{staples2021coupled}. Many works have been focused around the consequences of attenuation due to fluid viscosity \cite{zhu1995propagation, wu1995alternative,nayfeh1997excess,cegla2005material} on the modes of fluid-loaded plates. Nevertheless, the majority of these studies have been based around `hard' interfaces, presumably
since this is the regime most common in sensing and non-destructive applications.\ Recently in \cite{staples2021coupled} the case of water-loaded acrylic plates, which is a `soft' interface, was considered, and it was illustrated that significant differences arise in the phase speed of the coupled plate--Sc with respect to standard metal interfaces. In particular, they justified and experimentally demonstrated the dispersive behaviour of the symmetric coupled plate--Scholte mode which had previously only been characterized for soft films at very high frequencies \cite{xu2008effect}. Although not provided in their analysis, it is remarked in the experimental verification of \cite{staples2021coupled} the importance of viscoelastic properties of soft media, and in particular the importance of stress relaxation under particular frequency ranges. In this paper, we contribute to the theory in this study, by taking into account these dissipative mechanisms in order to assess their influence (and that of boundary layers) in both the symmetric and anti-symmetric coupled plate--Scholte modes for a wide range of plate thicknesses, and frequencies.




In Section \ref{Section:Governing equations}, we briefly introduce the visco-elastic (VE) formulation that we will employ throughout this work which allows us to simultaneously consider (visco)elastic solid and visco-acoustic fluid media as explained in \cite{garcia2022unified}. We discuss two classical models to capture the frequency dependence of the elastic moduli, namely the Kelvin--Voigt model (KVM) and the SLSM. These two models will be compared repeatedly in this work. In Section \ref{Section: Single Interface} we analyse the single interface configuration consisting of two half-space media in perfect contact. By seeking interface solutions to this set-up, we derive the \textit{Stoneley} DE, from which the \textit{Scholte} and \textit{Rayleigh} DEs can be obtained by considering the inviscid fluid and zero density limits respectively. The Stoneley and Scholte DEs give rise to Sc and LR waves which we discuss in detail. Since our fluid of interest is water, we study `hard' and `soft' interfaces (definitions are provided) by considering material parameters associated to steel and PVC, and investigate thermo-viscous effects. For the soft case, we analyse the additional behaviour that arises when the SLSM is employed.

In Section \ref{Section: Double interface general} we consider the double interface configuration, with the finite width medium constituting the slit/plate. We show how generalised DEs for symmetric/anti-symmetric modes valid for both configurations can be obtained. These equations recover more common forms found in the literature when certain effects are neglected such as the fluid's viscosity, which is illustrated. In the short wavelength-limit the general DEs reduce to the half-space Stoneley DE considered in Section \ref{Section: Single Interface}, which is used as the initial value of the recursive root finding technique. For the slits, we show how the softness of the solid, together with viscoelastic damping gives rise to rather different results to those presented for water-filled steel slits in \cite{cotterill2018thermo}, {and we confirm the claim made therein, namely that including thermal effects do not introduce any notable effects (within the considered regions of parameter space).}

For plates, both the symmetric/anti-symmetric coupled plate-Scholte modes are considered, and we observe the dispersive behaviour for the symmetric mode for soft plates reported recently in \cite{staples2021coupled}. We extend the theoretical analysis in \cite{staples2021coupled} by including the effect of thermo-viscous boundary layer attenuation (which is minimal) as well as viscoelastic damping in the solid which we show can be very important, especially near the \textit{glass transition} of the moduli. The presence of a global maximum of the attenuation as a function of plate thickness for the symmetric mode is highlighted. We finish with conclusions in Section \ref{section:conclusions}. {The general framework with the inclusion of thermal effects is devoted to Appendix \ref{appendix: TVE}, together with the equivalent TVE dispersion equations and a final note on the comparisons of TVE and VE solutions.}

\section{Governing equations for linear, isotropic VE continua}\label{Section:Governing equations}
We assume that the media under consideration here are linear, isotropic and further we make the approximation that all deformations are \textit{isothermal}, so that thermo-mechanical coupling need not be taken into account {(these effects are instead discussed in Appendix \ref{appendix: TVE})}. Consequently, the energy balance equation is automatically satisfied, so that the focus is to solve the linearised equation of motion, namely
\begin{equation}\label{eqn: of motion time domain}
    \nabla \cdot \hat{\bm{\sigma}}= \rho_0 \frac{\partial^2 \hat{\mathbf{u}}}{\partial t^2}, 
\end{equation}
where $\rho_0$ denotes the constant mass density, $\hat{\mathbf{u}}=\{\hat{u}_x,\hat{u}_y,\hat{u}_z\}$ the continuum's displacement vector, $\nabla$ is the vectorial gradient operator and  $\hat{\bm{\sigma}}$ is the Cauchy stress tensor which must capture all the required properties of the media we want to consider. Hereditary integrals give a general way to express the VE constitutive behaviour of the medium (e.g.\ \cite{christensen2012theory}, Chapter 1)
\begin{align}\label{eqn:stress time domain}
    \hat{\bm{\sigma}} = \int_{- \infty}^{{t}} 2 \hat{\mu}({t}-{\mathcal{T}})
    \frac{\partial \hat{\bm{e}}(\mathcal{T})}{\partial \mathcal{T}} \diff {\mathcal{T}} + \left(\int_{- \infty}^{{t}}  \hat{K}({t}-{\mathcal{T}}) \tr \left(\frac{\partial \hat{\bm{\varepsilon}}(\mathcal{T})}{\partial \mathcal{T}} \right) \diff {\mathcal{T}} \right) \bm{I}, 
\end{align}
where $\hat{\bm{\varepsilon}} = (\nabla \hat{\mathbf{u}} + (\nabla \hat{\mathbf{u}})^T )/2$ is the linearised strain tensor, $\tr{(\cdot)}$ the trace operator, $\bm{I}$ the identity tensor and $\hat{\bm{e}}=\hat{\bm{\varepsilon}} - \tr{(\hat{\bm{\varepsilon}})} \bm{I}/3$ represents the off-diagonal terms of the strain tensor. Further, $\hat{\mu}(t), \hat{K}(t)$ are time-dependent (visco) elastic moduli, which must be identically zero for $t<0$ to obey causality and are such that the integrals in (\ref{eqn:stress time domain}) are convergent. In this paper we will be considering time-harmonic disturbances, so that all the fields satisfy
\begin{equation}\label{eqns:TimeHarmonic}
   \{\hat{\mathbf{u}},\hat{\bm{\sigma}},\hat{\bm{\varepsilon}},\hat{\bm{e}} \} ({\mathbf{x},t)}= \operatorname{Re}{\{ \{\mathbf{u},\bm{\sigma},\bm{\varepsilon},\bm{e}\}({\mathbf{x})}\me^{-\mi {\omega} {t}}\}},
\end{equation}
which when substituted in (\ref{eqn:stress time domain}), (\ref{eqn: of motion time domain}) and factoring out the common term give respectively
\begin{subequations}\label{VE eqnsss}
\begin{align} 
\label{eqn:timeharmonicstressVEE}
    & {{\bm{\sigma}}} = 2{{\mu}}(\omega) {\bm{e}} + {{K}}( {\omega}) \tr({\bm{\varepsilon}})\bm{I},\\
\label{eqn:timeharmonicmomentumVEE}
    ({{K}}( {\omega}) +& \frac{4}{3}{{\mu}}(\omega)) {\mathbf{\nabla}} \left({\mathbf{\nabla}} \cdot {{\mathbf{u}}} \right) - {{\mu}}(\omega) {\mathbf{\nabla}} \times {\mathbf{\nabla}} \times  {{\mathbf{u}}} +{\rho}_0 {\omega}^2 {{\mathbf{u}}} = \mathbf{0},
\end{align}
\end{subequations}
where $\mu(\omega), K(\omega)$ are scaled Fourier transforms of the original moduli \cite{garcia2022unified}. On adopting the Helmholtz decomposition
\begin{equation}
    \label{viscoelastic helmholtz decomposition}
{{\mathbf{u}}} = {\nabla} {\phi} + {\nabla} \times {\mathbf{\Phi}}, 
\end{equation}
with $\nabla \cdot {\mathbf{\Phi}}=0$, the compressional and shear wave potentials $\phi,\mathbf{\Phi}$ must respectively satisfy
\begin{subequations} \label{viscoelastic phi&Phi eqn}
\begin{align} \label{viscoelastic phi eqn}
    &\left(\Delta  +   k_\phi^2\right){\phi} = 0, \qquad &&k_\phi^2(\omega) = \frac{{\rho}_0 {\omega}^2}{{{K}}( {\omega}) + \frac{4}{3}{{\mu}}(\omega)},\\
    &\left(\Delta  + k_\Phi^2\right){\mathbf{\Phi}} = \mathbf{0}, \qquad && k_\Phi^2(\omega) = \frac{{\rho}_0 {\omega}^2}{{{\mu}}(\omega)} \label{viscoelastic Phi eqn},
\end{align}
\end{subequations}
recalling that the material parameters appearing in the compressional/shear wavenumbers $k_\phi, k_\Phi$ in (\ref{viscoelastic phi&Phi eqn}) are isothermal, which is particularly important to consider for viscous gases such as air (since the isothermal bulk modulus can differ significantly from the corresponding more common adiabatic modulus) \cite{garcia2022unified}. Given the shear and bulk moduli present in (\ref{viscoelastic phi&Phi eqn}), we can define the generalized first Lam\'e coefficient, Poisson's ratio and Young's modulus respectively as \cite{tschoegl2002poisson}
\begin{equation}\label{eqn:nonlocalTVE lambda}
    {{\lambda}}(\omega) = {{K}}(\omega) - \frac{2}{3}{{\mu}}(\omega) ,
\qquad  {{\nu}}(\omega) = \frac{3{{K}}(\omega) - 2 {{\mu}}(\omega)}{6 {{K}}(\omega) + 2 {{\mu}}(\omega)},\qquad  {{E}}(\omega) = \frac{9 {K}(\omega) {{\mu}}(\omega) }{3{{K}}(\omega) +  {{\mu}}(\omega)}. 
\end{equation}
There exist plenty of models to capture the frequency dependence of these moduli which are appropriate in particular circumstances (e.g.\ \cite{borcherdt2009viscoelastic}), but in this work we focus on the Kelvin--Voigt Model (KVM) (which we also refer to as local VE) and the Standard Linear Solid Model (SLSM), which are discussed extensively in \cite{garcia2022unified} starting from their time domain behaviour. For a given modulus $\mathcal{M}(\omega)$ we have
\begin{subequations}\label{eqn: 1D VE}
\begin{align}\label{eqn: 1D KV}
   (\text{KVM}) \quad \mathcal{M}(\omega) &= \mathcal{M}_0 - \mi \omega \eta_\mathcal{M}, \\ \label{eqn: 1D SLSM}
   (\text{SLSM}) \quad \mathcal{M}(\omega) &= \mathcal{M}_\infty - (\mathcal{M}_0 - \mathcal{M}_\infty ) \frac{\mi \omega t_r }{1 - \mi \omega t_r}=\frac{\mathcal{M}_\infty+\mathcal{M}_0(\omega t_r)^2}{1+(\omega t_r)^2}-\mi \frac{(\mathcal{M}_0-\mathcal{M}_\infty)\omega t_r}{1+(\omega t_r)^2}.
\end{align}
\end{subequations}
In the KVM, the real part ($\mathcal{M}_0$) is fixed and the (attenuative)  imaginary part known as the \textit{loss modulus} is characterized by $\omega \eta_\mathcal{M}$ where $\eta_\mathcal{M}$ is a constant viscosity coefficient. In the SLSM, $\mathcal{M}_\infty, \mathcal{M}_0$ correspond to the long-term and instantaneous moduli and $t_r$ represents the (single) relaxation time of the material. The {Deborah number} $\omega t_r$ in (\ref{eqn: 1D SLSM}), relates the characteristic time of the problem ($1/\omega$) with the relaxation time of the material in consideration. The moduli further satisfy $\mathcal{M}_\infty \leq \mathcal{M}_0$, with the equality implying the medium is perfectly elastic. We note that for $\omega t_r \gg 1$, $\mathcal{M}\rightarrow \mathcal{M}_0$ and conversely for $\omega t_r \ll 1$, $\mathcal{M}\rightarrow \mathcal{M}_\infty$ and therefore higher frequencies correspond to the \textit{glassy} (stiffer) phase of the materials and lower frequencies to a \textit{rubbery} (softer) phase. The loss modulus has a global maximum at $\omega t_r =1$ which defines the \textit{glass transition region}, and therefore it is in the vicinity of this region where the majority of intrinsic VE losses are manifested.\\

Visco-acoustic (Newtonian) fluids can also be described by (\ref{VE eqnsss})-(\ref{viscoelastic phi&Phi eqn}) with ${\mu}(\omega) = -\mi \omega \eta_\mu$, where  $\eta_\mu>0$ is the kinematic viscosity \cite{garcia2022unified, wu1995alternative, cegla2005material}, so that the square of the shear wavenumber becomes purely imaginary, i.e.\ $k_{\Phi}^2= \mi/ {\delta}^2_\nu$ where ${\delta}_\nu = \sqrt{\eta_\mu/\rho_0 \omega}$ is a boundary layer parameter related to the common Stokes' boundary layer thickness by ${\delta}_s=2\pi \sqrt{2} {\delta}_\nu$.

\section{Single interface: Two VE half-spaces in perfect contact} \label{Section: Single Interface}
\begin{figure}[H]
\centering 
\def\svgwidth{\columnwidth}
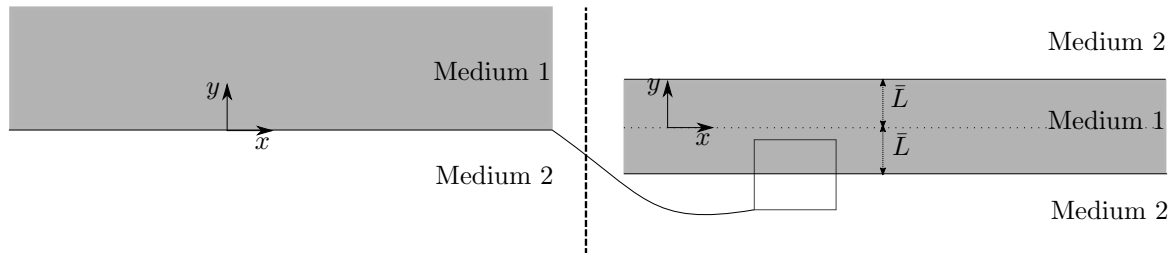
\caption{Schematic representations of the 2D geometric configurations considered in this paper, noting that all media are infinitely extending in the $x$-direction. Left: Two semi-infinite VE continua as considered in Section \ref{Section: Single Interface} (single interface). Right: A VE medium of width $\Bar{W} = 2\Bar{L}$ bounded by two semi-infinite VE media as considered in Section \ref{Section: Double interface general} (double interface).}
\label{Fig:Geometries}
\end{figure}

We first seek interface waves associated with the configuration of VE-VE half-spaces separated by an interface at $y=0$, as seen on the left of Figure \ref{Fig:Geometries}. From the resulting DEs we can determine the Leaky Rayleigh and Scholte--Stoneley modes and analyse some of their key properties. As we will observe in the next section, the understanding of the half-space will facilitate the study of the double interface configuration (right of Figure \ref{Fig:Geometries}).

\subsection{The Stoneley, Rayleigh and Scholte Dispersion Equations}

 In Section \ref{Section:Governing equations} all the relevant equations were given in dimensional form. It will be convenient in the subsequent analysis to have a notational distinction between dimensional/non-dimensional quantities and therefore from here onwards we will introduce an over-bar $\bar{\cdot}$ to all dimensional quantities. We seek potential solutions to the governing equations (\ref{VE eqnsss}) that decay away from the boundary in each half-space, since interface waves can be represented as a linear combination of these four waves:

\begin{subequations}\label{Potentials:Stoneley}
\begin{align} \label{Stoneley Compressional}
\bar{\phi}_1 &= \bar{P}_1 \me^{ - \bar{\gamma}_{\phi_1} \bar{y} + \mi \bar{k} \bar{x}},  
& \bar{\phi}_2 &= \bar{P}_2\me^{  \bar{\gamma}_{\phi_2} \bar{y} + \mi \bar{k} \bar{x}},\\ 
\label{Stoneley Shear}
\bar{\Phi}_1 &= \bar{S}_1 \me^{ - \bar{\gamma}_{\Phi_1} \bar{y} + \mi \bar{k} \bar{x}},  &
\bar{\Phi}_2 &= \bar{S}_2 \me^{ \bar{\gamma}_{\Phi_2} \bar{y} + \mi \bar{k} \bar{x}}, 
\end{align}
\end{subequations}
where 
\begin{equation}\label{defn: sqrt functions HS}
    \bar{\gamma}_{\phi_1} = (\bar{k}^2 - \bar{k}_{\phi_1}^2)^{1/2}, \quad \bar{\gamma}_{\phi_2} = (\bar{k}^2 - \bar{k}_{\phi_2}^2)^{1/2}, \quad \bar{\gamma}_{\Phi_1} = (\bar{k}^2 - \bar{k}_{\Phi_1}^2)^{1/2}, \quad \bar{\gamma}_{\Phi_2} = (\bar{k}^2 - \bar{k}_{\Phi_2}^2)^{1/2},
\end{equation}
 for some complex amplitudes $\bar{P}_1,\bar{P}_2,\bar{S}_1,\bar{S}_2$. We must ensure that the choice of the various branch cuts of the square root functions in (\ref{defn: sqrt functions HS}) is consistent with causality, which is further discussed below. At the interface between the two media, the solutions must satisfy continuity of traction and displacement boundary conditions (BCs), that is on $\bar{y}=0$,
\begin{equation}\label{VE 2 HS BCs 1}
    \bar{\sigma}_{yy}^1 = \bar{\sigma}_{yy}^2, \;\;  
    \bar{\sigma}_{xy}^1 = \bar{\sigma}_{xy}^2, \;\;
    \bar{\mathbf{u}}_1 = \bar{\mathbf{u}}_2.
\end{equation}
Substitution of (\ref{Potentials:Stoneley}) into (\ref{VE 2 HS BCs 1}) then leads to the VE Stoneley DE, that is
\begin{equation}
\begin{aligned}\label{StoneleyDE}
    \bar{\mathcal{D}}_\text{St} = \bar{c}^4 [( \bar{\rho}_1-\bar{\rho}_2)^2-&( \bar{\rho}_2{A_1}+ \bar{\rho}_1{A_2})
   (\bar{\rho}_2{B_1} + \bar{\rho}_1{B_2} )] \\
   & +2 \bar{c}^2 \bar{\mathcal{Q}} (-\bar{\rho}_2{A_1} {B_1}
   + \bar{\rho}_1{A_2} {B_2}-\bar{\rho}_1+\bar{\rho}_2) + \bar{\mathcal{Q}}^2
   ({A_1} {B_1}-1) ({A_2} {B_2}-1) = 0,
   \end{aligned}
\end{equation}
which for a purely elastic--elastic interface takes the form as in Stoneley's original paper \cite{stoneley1924elastic}, (as expected prior to the specification of the frequency dependence of the elastic moduli due to the correspondence principle for elasticity see e.g.\ \cite{borcherdt2009viscoelastic}) with 
\begin{equation}\label{definition: HS quantities}
    \bar{c}=\frac{\bar{\omega}}{\bar{k}}, \qquad \{A_1, A_2, B_1, B_2 \} = \frac{1}{\bar{k}}\{\bar{\gamma}_{\phi_1}, \bar{\gamma}_{\phi_2}, \bar{\gamma}_{\Phi_1}, \bar{\gamma}_{\Phi_2} \},  \qquad \bar{\mathcal{Q}} = 2(\bar{\rho}_1 \bar{c}_{\Phi_1}^2 - \bar{\rho}_2 \bar{c}_{\Phi_2}^2),
\end{equation}
 where $\bar{c}_{\Phi} =\bar{\omega}/\bar{k}_{\Phi}$ in each medium. { When thermal effects are considered, (\ref{StoneleyDE}) takes the more complex form (\ref{stoneley:TVE-TVE}), as discussed in Appendix \ref{appendix:section: TVE HS}}. If we let the density of the upper medium vanish, i.e.\ $\bar{\rho}_1 \rightarrow 0$ in (\ref{StoneleyDE}), after some manipulation using (\ref{definition: HS quantities}) we obtain the Rayleigh DE, namely
\begin{equation}\label{RayleighDE}
    \bar{\mathcal{D}}_\text{Ra} = (2\bar{k}^2 - \bar{k}_{\Phi_2}^2)^2 - 4 \bar{k}^2 \bar{\gamma}_{\phi_2} \bar{\gamma}_{\Phi_2} = 0,
\end{equation}
which governs the ubiquitous \textit{Rayleigh waves}, originally described in \cite{rayleigh1885waves} for elastic media. Alternatively, in the inviscid fluid limit $\bar{\mu}_1(\omega) \rightarrow 0$, it follows that $\bar{k}_{\Phi_1}^2 \rightarrow \infty$ and hence $\bar{\gamma}_{\Phi_1}, B_1 \rightarrow \infty$ whereas $\bar{\mathcal{Q}}\rightarrow -2  \bar{\rho}_2 \bar{c}_{\Phi_2}^2$ so that (\ref{StoneleyDE}) becomes
\begin{equation}
     \label{ScholteDE}
      \bar{\mathcal{D}}_\text{Sc} = \bar{\mathcal{D}}_\text{Ra} + 
     \frac{\bar{\gamma}_{\phi_2} \bar{k}_{\Phi_2}^4}{\bar{\gamma}_{\phi_1} \rho_s}=0, \quad \text{with} \quad  \rho_s = \frac{\bar{\rho}_2}{\bar{\rho}_1}, 
\end{equation}
which is the common Scholte DE (also referred to as Scholte--Stoneley DE \cite{staples2021coupled}). As we will discuss shortly, in general (\ref{ScholteDE}) and (\ref{StoneleyDE}) admit two families of interface wave solutions, namely \textit{Leaky Rayleigh} and \textit{Scholte} waves. Direct observation of (\ref{ScholteDE}) illustrates the role of the density ratio $\rho_s$, when $\rho_s \gg 1$ i.e.\ for ``hard" interfaces, the fluid-loading term will have negligible influence and therefore the behaviour of $\bar{\mathcal{D}}_\text{Sc}$ can be seen as a small perturbation to $\bar{\mathcal{D}}_\text{Ra}$. Nevertheless, for softer interfaces such that $\rho_s \approx 1$ significantly different behaviour of the spectrum can be expected.

Naturally, due to the symmetry of the configuration,  (\ref{RayleighDE}) and (\ref{ScholteDE}) can also be recovered by taking the limits $\bar{\rho}_2 \rightarrow 0$, $\bar{\mu}_2(\omega) \rightarrow 0$ respectively (and interchanging the subscripts `$1$' and `$2$' in the subsequent equations). Despite the interest of this work being on losses in fluid--solid configurations, this symmetry property makes the use of the general VE--VE configuration convenient, which will become apparent in Section \ref{Section: Double interface general} where we will be able to consider two physically different problems with the same set of DEs.\\

The roots of (\ref{StoneleyDE}), (\ref{RayleighDE}), (\ref{ScholteDE}), (\ref{stoneley:TVE-TVE}) and the various DEs studied in Section \ref{Section: Double interface general} are calculated using the MATLAB [version 9.8.0.1380330 (R2020a)] command \textit{fsolve}, which finds the local zero of a function close to a given starting point specified by the user. This initial value will differ depending on the mode under consideration, as we will specify below.\\ 

Throughout the paper, the several square root functions (\ref{defn: sqrt functions HS}) are chosen such that, as $|\bar{k}| \rightarrow \infty$ $\bar{\gamma}_{\phi_1}, \bar{\gamma}_{\phi_2}, \bar{\gamma}_{\Phi_1}, \bar{\gamma}_{\Phi_2} \rightarrow \Bar{k}$, with the branch cuts from $\bar{k}_{\phi_1}, \bar{k}_{\phi_2}, \bar{k}_{\Phi_1}, \bar{k}_{\Phi_2}$ taken in the upper half-plane and those from $-\bar{k}_{\phi_1}, -\bar{k}_{\phi_2}, -\bar{k}_{\Phi_1}, -\bar{k}_{\Phi_2}$ taken in the lower half-plane. The branch cuts are chosen to run with fixed real parts (i.e.\ parallel to the imaginary axis) from the respective branch points. We note that in order to analyse \textit{all} roots of these equations, in principle it is necessary to consider all possible combinations of Riemann sheets giving e.g.\ $2^4$ for (\ref{StoneleyDE}), $2^3$ for (\ref{ScholteDE}), $2^2$ for (\ref{RayleighDE}) and  $2^6$ for (\ref{stoneley:TVE-TVE}), see e.g.\ \cite{mozhaev2002subsonic, schroder2001complex, harris2002comment}. In this work, we are concerned with the influence of TVE effects on the well established roots of these equations, so we simply need to make sure the obtained solutions are causal and do not jump Riemann sheets. On top of this, we ensure they behave as expected in the absence of any form dissipation.\\

In what follows, the roots of the DEs will be plotted in terms of non-dimensional phase speed and attenuation, which are given by
\begin{subequations}
\begin{align}
   \text{Phase Speed } &= \frac{\bar{v}}{\bar{c}_\square} = \frac{\operatorname{Re} \{\bar{c} \}}{\bar{c}_\square} = \frac{\bar{\omega}}{\operatorname{Re} \{\bar{k}\} \bar{c}_\square}, \\ \label{definition: attenuation p wavelength} \text{Attenuation (dB/wavelength)} &=40 \pi \frac{\operatorname{Im}\{\bar{k}\}}{\operatorname{Re}\{\bar{k}\}} \log_{10}(\me),
\end{align}
\end{subequations}
where $\bar{c}_\square$ represents a particular sound velocity which we assume constant. In particular $\bar{c}_0$ denotes the speed of sound of water, and $\bar{c}_s$ is the lossless shear wave speed of sound in a certain solid material (from Table \ref{table: VE parameters} water gives $\bar{c}_0=1490$ m/s, steel $\bar{c}_s=3000$ m/s, and PVC $\bar{c}_s=1100$ m/s).

\subsection{Material parameters: hard/soft solids}\label{subsection:soft/hard solids}
We have seen above that different limits of the density ratio $\rho_s$ can give some insights on the importance of certain terms in the various DEs above. Generally, knowing whether a given fluid--solid pair is hard/soft is intuitive although in some instances this can lead to ambiguity, as recently noticed in \cite{staples2021coupled}, which is focused on the Scholte mode solution to (\ref{ScholteDE}). Indeed, a common definition is whether the material parameters are such that $\operatorname{Re}\{\bar{k}_\phi\} < \operatorname{Re}\{\bar{k}_\Phi\} \leq \operatorname{Re}\{\bar{k}_\text{F} \}$ or $\operatorname{Re}\{\bar{k}_\phi \} \leq \operatorname{Re}\{\bar{k}_\text{F}\} < \operatorname{Re}\{\bar{k}_\Phi\}$ which correspond to hard or soft respectively, as in e.g.\ \cite{glorieux2001character}. Note that since we are fixing materials 1/2 to be fluid/solid respectively (for definition purposes) we have written $\bar{k}_\text{F} \equiv \bar{k}_{\phi_1}$ for the visco-\textit{acoustic} wavenumber, and $\bar{k}_\phi \equiv \bar{k}_{\phi_2}$, $\bar{k}_\Phi \equiv \bar{k}_{\Phi_2}$ corresponding to the standard pressure and shear VE wavenumbers. Nevertheless, in some instances the definition implied by the inequalities can be physically inaccurate e.g.\ it is shown in \cite{staples2021coupled} by considering the transition between $\operatorname{Re}\{\bar{k}_\Phi\}$, and $\operatorname{Re}\{\bar{k}_\text{F}\}$ that the Scholte mode's phase speed remains constant (which should not be the case when transitioning between a `hard' and `soft' solid). For this reason, in that work a hard interface is defined as one where the Scholte velocity is approximately equal to the speed of sound in the fluid, and conversely a soft interface is one where the Scholte velocity is notably less than the speed of sound in the fluid. As a result, the latter definition is more inspired by the actual physics, whereas the former indicated by the inequalities above is purely motivated by the complex plane spectrum. In \cite{zhu2004leaky} the same idea is discussed in terms of acoustic impedance of the fluid/solid pair.

\begin{table}[H]
\centering
\small
\begin{tabular}{ |p{5.3cm}||p{1cm}|p{1.2cm}|p{1.2cm}|p{1.3cm}|p{1.1cm}|p{1.1cm}|}
 \hline
 \multicolumn{7}{|c|}{ {\textbf{VE Parameter Values (Kelvin--Voigt Model)}}} \\
 \hline
 \hspace{1.5cm} Parameter& Unit & Symbol  & Water & Steel & PVC & $M_{\tau=0.9}$ \\
 \hline
 Background density &   kg/m$^3$  & ${\rho}_0$  & 1000 & 7871 & 1360 & 2011\\ 
 Shear modulus &   GPa  & ${\mu}_0$   & -- & 70.839 &  1.65 & 8.55 \\ 
 Bulk modulus (Adiabatic) &  GPa  & $K_0$  & 2.2222  & 189.54 &  5.0415 & 24.342\\ 
 Young's modulus (Adiabatic)&   GPa  & ${E}_0$   & -- &   & 4.4524 & \\ 
 Dynamic shear viscosity & Pa$\cdot$s & ${\eta}_\mu$ & $ 10^{-3}$ & 1.85$\times 10^{-5}$ & $0.1$& 0.09\\ 
 Dynamic bulk viscosity & Pa$\cdot$s & ${\eta}_K$ & 3$\times 10^{-3}$ & 1.1$\times 10^{-5}$ & $0.3$& 0.27 \\
 \hline
\end{tabular}
\caption{Parameter values for water, steel and PVC used in calculations, taken from \cite{pierce1981acoustics}, \cite{cotterill2018thermo} and \cite{favretto1996theoretical} respectively, assumed to be independent of frequency. $M_{\tau=0.9}$ is defined in (\ref{linear transition steel->PVC}). The PVC values are only applicable for local VE (KVM); the values for the stress relaxation discussion are given in Sections \ref{subsection: HS stress relaxation}, \ref{subsection: Slit stress relaxation} and \ref{subsection:Plates relaxation}. The additional thermal parameters required for the TVE calculations are provided in Table \ref{table: Th parameters}.}
\label{table: VE parameters}
\end{table}

Our primary physical interest here is in water--solid interfaces, so that in most of what follows medium 1 in (\ref{Potentials:Stoneley}) is fixed by the parameters of water, which are given in Table \ref{table: VE parameters}. For a given fluid, the behaviour of the interface modes can be greatly influenced by the (visco-)elastic properties of the solid, which we explore below. As is done in several works (e.g.\ \cite{glorieux2001character}) in order to illustrate this, we will concentrate on two solids of opposing nature, namely steel and PVC whose values are also shown in Table \ref{table: VE parameters}. As we will showcase below, when compared to water these values correspond to hard/soft materials respectively (for both definitions presented above). Furthermore, in order to aid our study of the effect for the hard-soft transition on the relevant modes, we will make use of a linear continuous transition in all Material 2 parameters (listed in Tables \ref{table: VE parameters}, \ref{table: Th parameters}) through a function $\mathbf{f}:[0,1] \rightarrow \mathbb{R}^n$ s.t.
\begin{equation}\label{linear transition steel->PVC}
    \mathbf{f}(\tau) = [\textbf{Steel}](1-\tau) + [\textbf{PVC}] (\tau) \qquad 0 \leq \tau \leq 1,
\end{equation}
where $[\textbf{Steel}], [\textbf{PVC}]$ are simply vectors encoding all the physical parameters of each material. With regards to the frequency dependence of the VE moduli in the solid, we start by considering the KVM in Sections \ref{subsubsec: LR Mode}, \ref{subsubsec: Sch Mode} following (\ref{eqn: 1D KV}) for both the shear and bulk moduli, characterized by the viscosity coefficients $\eta_\mu ,\eta_K$ respectively. Consequently, the SLSM from (\ref{eqn: 1D KV}) is employed in Section \ref{subsection: HS stress relaxation}, with further details provided there. { For convenience, thermal effects are only illustrated with the KVM.}

\subsection{Modes at the interface of 2 semi-infinite half-spaces}
Having defined the relevant equations to our physical set-up and material parameters in consideration, we now focus our attention to the solutions of the Stoneley DE (\ref{StoneleyDE}), that can give rise to Leaky--Rayleigh and Stoneley--Scholte modes, which are considered individually next.

\subsubsection{The Leaky--Rayleigh (LR) Mode}\label{subsubsec: LR Mode}
We denote the LR mode solutions to (\ref{StoneleyDE}), (\ref{ScholteDE}) by $\bar{k}_\text{LR}=\bar{\omega}/\bar{c}_\text{LR}$. It is well established that the LR wave propagates slightly slower than the associated solid's shear body wave, marginally faster than the ordinary Rayleigh wave and attenuates in the direction of propagation due to part of the energy being shed into acoustic waves (radiated) into the fluid  \cite{zhu2004leaky}. In most instances, this leakage is a consequence of the LR being \textit{supersonic} ($\operatorname{Re}\{\bar{k}_\text{LR}\} <  \operatorname{Re}\{\bar{k}_\text{F}\}$). In fact, this was believed to be necessary for its existence (see e.g.\ \cite{glorieux2001character}) prior to the work of \cite{mozhaev2002subsonic}, who showed the existence of subsonic leakage (although in a very small region) by careful analysis of the complex plane spectrum.
Nevertheless, in most circumstances the LR cannot propagate and the inequality ($\operatorname{Re}\{\bar{k}_\Phi\} < \operatorname{Re}\{\bar{k}_\text{LR}\} \leq \operatorname{Re}\{\bar{k}_\text{F}\}$) usually holds. In practical conditions, the LR mode is also subject to an extra attenuating mechanism, namely that of visco-thermal boundary layer effects in the fluid, as well as the TVE damping within the solid. In \cite{wu1995alternative} it was explicitly found that viscous effects dominate over heat conduction effects (especially for water) based on the original work \cite{qi1994attenuated}, who first concluded that the effect of viscosity can be neglected for fluids with Reynolds number larger than 2500. We next showcase these differences by comparing solutions between the VE Stoneley DE, (\ref{StoneleyDE}) with the inviscid fluid equivalent solution (\ref{ScholteDE}) as well as the TVE Stoneley DE (\ref{stoneley:TVE-TVE}). 

Given the parameters in Table \ref{table: VE parameters} and the discussion above, we note that for a water--PVC interface ($\tau = 1 $ in (\ref{linear transition steel->PVC})) the LR cannot be supported, since the shear wave is highly subsonic ($ \operatorname{Re}\{\bar{k}_\text{F}\} < \operatorname{Re}\{\bar{k}_\Phi\}$), whereas for water--Steel ($\tau = 0$ in (\ref{linear transition steel->PVC})) the mode will exist. Starting from $\tau=0$, as we transition through increasing $\tau$ we observe the LR root becomes increasingly attenuative (per unit metre), until it reaches the subsonic region. A detailed analysis of the transition from the existence of the LR in the complex plane was provided in \cite{mozhaev2002subsonic} via an isotropic gold-silver alloy with variable content in contact with water.

\begin{figure}
    \centering
    \includegraphics[trim={1.5cm 0cm 0 5cm}, scale=0.6]{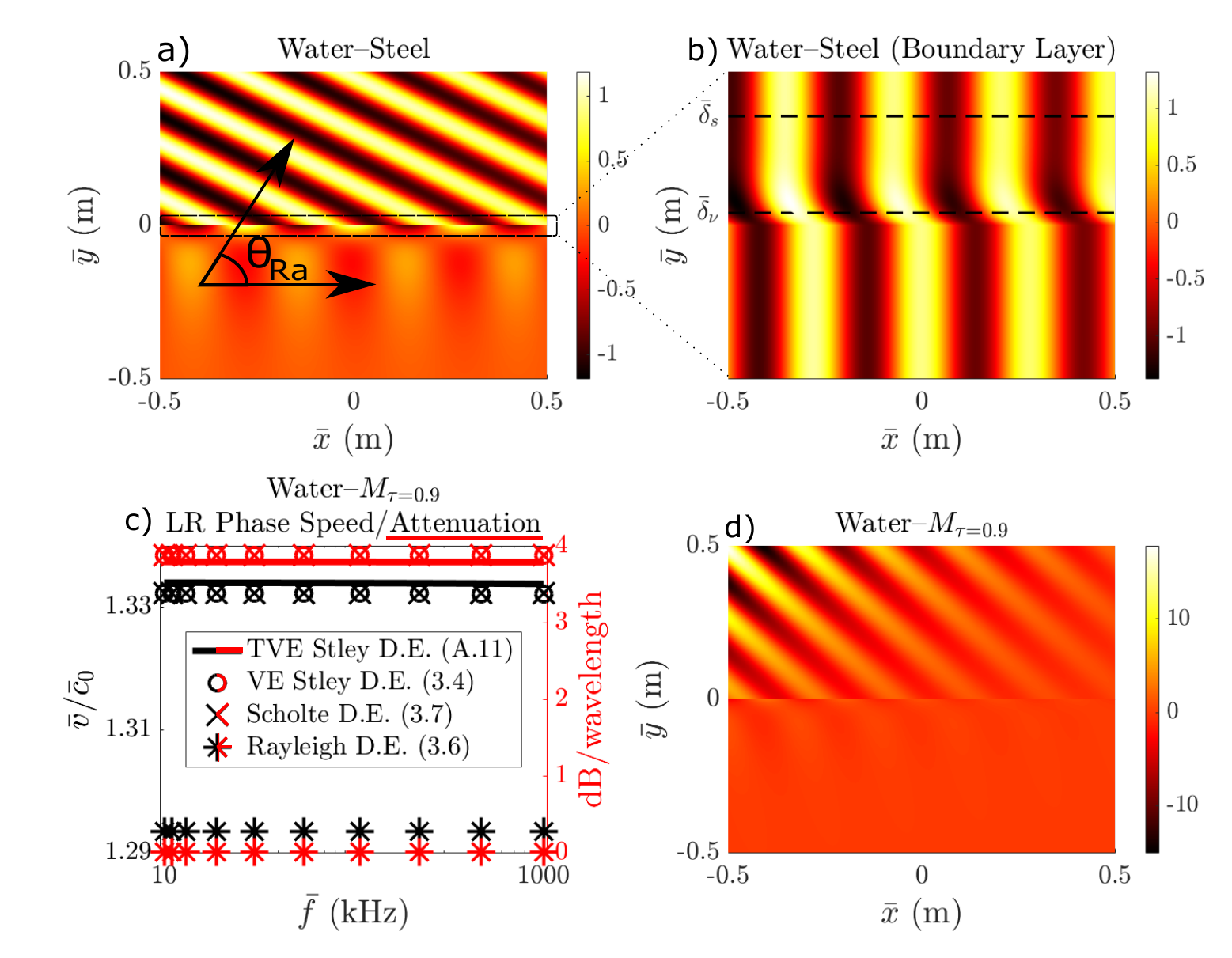}
    \caption{Heatmaps a), b), d) illustrate the dimensionless horizontal particle displacement $\operatorname{Re}\{u_x\}$ of the Leaky--Rayleigh mode from (\ref{StoneleyDE}) at $10$ kHz for water--steel (a), (b) and water--$M_{\tau=0.9}$ (d). (b) corresponds to the same calculation as (a) but near the boundary (as indicated by the small dotted lines) and all three plots have been normalized such that $u_x(y=x=0)=1$. In c) we give comparisons (in phase speed/attenuation) between the LR solutions obtained by different DEs for water--$M_{\tau=0.9}$, noting the two different $y$ scales represented by black/red.}
    \label{Fig:LeakyRayleigh}
\end{figure}

In order to numerically obtain the LR root using \textit{fsolve}, as an initial guess we simply choose a real value that is slightly slower than the solid's shear wave speed. Indeed, we obtain that the LR mode solution to (\ref{StoneleyDE}) for Steel at $10$ kHz gives $\bar{k}_\text{LR} =22.44 + 0.24 \mi$ m$^{-1}$, whereas for $M_{\tau=0.9}$ (the artificial material corresponding to $\tau = 0.9$ in (\ref{linear transition steel->PVC})), $\bar{k}_\text{LR} = 31.49 + 2.238 \mi$ m$^{-1}$, whose associated (normalized) horizontal particle displacements $\operatorname{Re}\{u_x\}$ are given in Figure \ref{Fig:LeakyRayleigh} a),b),d) (all of which have been normalized such that $u_x(y=x=0)=1$). We observe that the motion (and hence energy) of the LR is radiated as a pressure wave in the fluid propagating in the direction of the Rayleigh angle $\theta_\text{Ra}=\arctan \{\operatorname{Re}(\mi \bar{\gamma}_{\phi_1})/\operatorname{Re}(\Bar{k}_\text{LR}) \}$ which is measured anti-clockwise from $x=0$ (see Fig \ref{Fig:LeakyRayleigh}a)). For instance, water--steel gives $\theta_\text{Ra}=57.826^\circ$, whereas for water--$M_{\tau=0.9}=41.857^\circ$.
From the expanded Fig \ref{Fig:LeakyRayleigh}b) we observe the viscous boundary layer region in the fluid, which we remark is very thin (at $10$ kHz we have approximately $\bar{\delta}_\nu \approx 4$ $\mu$m). As with the traditional Rayleigh mode, the motion within the solid half-space is very localized near the boundary. The dissipation for increasing $x$ parallel to $y=0$ is very apparent for $M_{\tau=0.9}$, where the motion in the solid in Fig \ref{Fig:LeakyRayleigh}d) cannot be appreciated due to the higher magnitude of the mode along the positive $y$ direction (fluid region), as indicated by the colour bar.\\ 

When performing the same calculations as above but with the Scholte DE (\ref{ScholteDE}) (i.e.\ considering the inviscid fluid limit) we obtain very accurate answers for the phase speed and attenuation for both materials, despite the boundary layer effects observed for $u_x$ naturally not being captured since in this case we are allowing the fluid to slip in the horizontal direction. On the other hand, the (VE) pure Rayleigh DE (\ref{RayleighDE}) in the absence of fluid loading predicts the phase velocity fairly accurately, but fails to account for the attenuation\footnote{Particularly for $M_{\tau=0.9}$, which is in accordance with the observation noted just below (\ref{ScholteDE}) regarding the role of the density ratio $\rho_s$.}, as shown in Fig \ref{Fig:LeakyRayleigh}c), from which we can conclude that energy radiation is indeed the predominant effect contributing to the attenuation of this mode, with boundary layers and the solid's VE damping playing a much smaller role,  which would be inconsequential in most practical scenarios. {The result of thermal effects on the LR mode is included in Fig \ref{Fig:LeakyRayleigh}c), which we note gives a slightly higher phase speed as opposed to the isothermal equivalent.} More generally, in Fig \ref{Fig:LeakyRayleigh}c) we also observe almost no dispersion (at the $y$ scales provided) of the LR mode for $M_{\tau=0.9}$ which we also found to be the case for steel, and tested it up to MHz frequencies.

\subsubsection{The Scholte--Stoneley (Sc) Mode}\label{subsubsec: Sch Mode}
We denote the Sc mode solutions to (\ref{StoneleyDE}), (\ref{ScholteDE}), (\ref{stoneley:TVE-TVE}) by $\bar{k}_\text{Sc}=\bar{\omega}/\bar{c}_\text{Sc}.$ The Stoneley--Scholte mode is an acoustic surface wave whose velocity is \textit{subsonic} and slower than the bulk waves of the solid (i.e.\ $\operatorname{Re}\{\bar{k}_\phi \},\operatorname{Re}\{\bar{k}_\Phi \},\operatorname{Re}\{\bar{k}_\text{F}\} \leq \operatorname{Re}\{\bar{k}_\text{Sc}\}$) and, when neglecting the thermo-viscous effects of both liquid and solid, it travels unattenuated along the interface, and decays exponentially away from it. Unlike the LR case above, it is present for all fluid--solid interfaces, although its behaviour is highly influenced by the material properties in question, as pointed out in \cite{staples2021coupled}. Following Section \ref{subsection:soft/hard solids}, for a fixed fluid (in our case water), \textit{soft} solids yield deeper penetration depths of the Sc mode and are therefore more convenient for applications \cite{glorieux2001character}. With regards to losses in the fluid region, Stoneley--Scholte modes can be attenuated through two mechanisms, namely through boundary layer leakage of shear/vortical waves from the interface into the fluid, and through the longitudinal bulk waves in the thermo-viscous fluid. Nevertheless, for the latter to be noticeable the frequencies must be very high, and in general for both mechanisms to become important the viscosity of these fluids must be fairly high (e.g.\ in the experiments of \cite{cegla2005material} glycerol and honey are used). Indeed, \cite{volkenshtein1988structure}, \cite{gusev2015effect} concluded that for low viscosity fluids (including water), the effects on the Sc mode are such that it can generally be ignored. The VE effects within softer solids were studied theoretically and experimentally by Favretto-Anr\`es \cite{favretto1996theoretical}, \cite{favretto1997excitation} who found weak dispersion in the lower frequency range for synthetic resins in contact with water.\\

\begin{figure}
    \centering
    \includegraphics[trim={1.5cm 0cm 0 5cm}, scale=0.6]{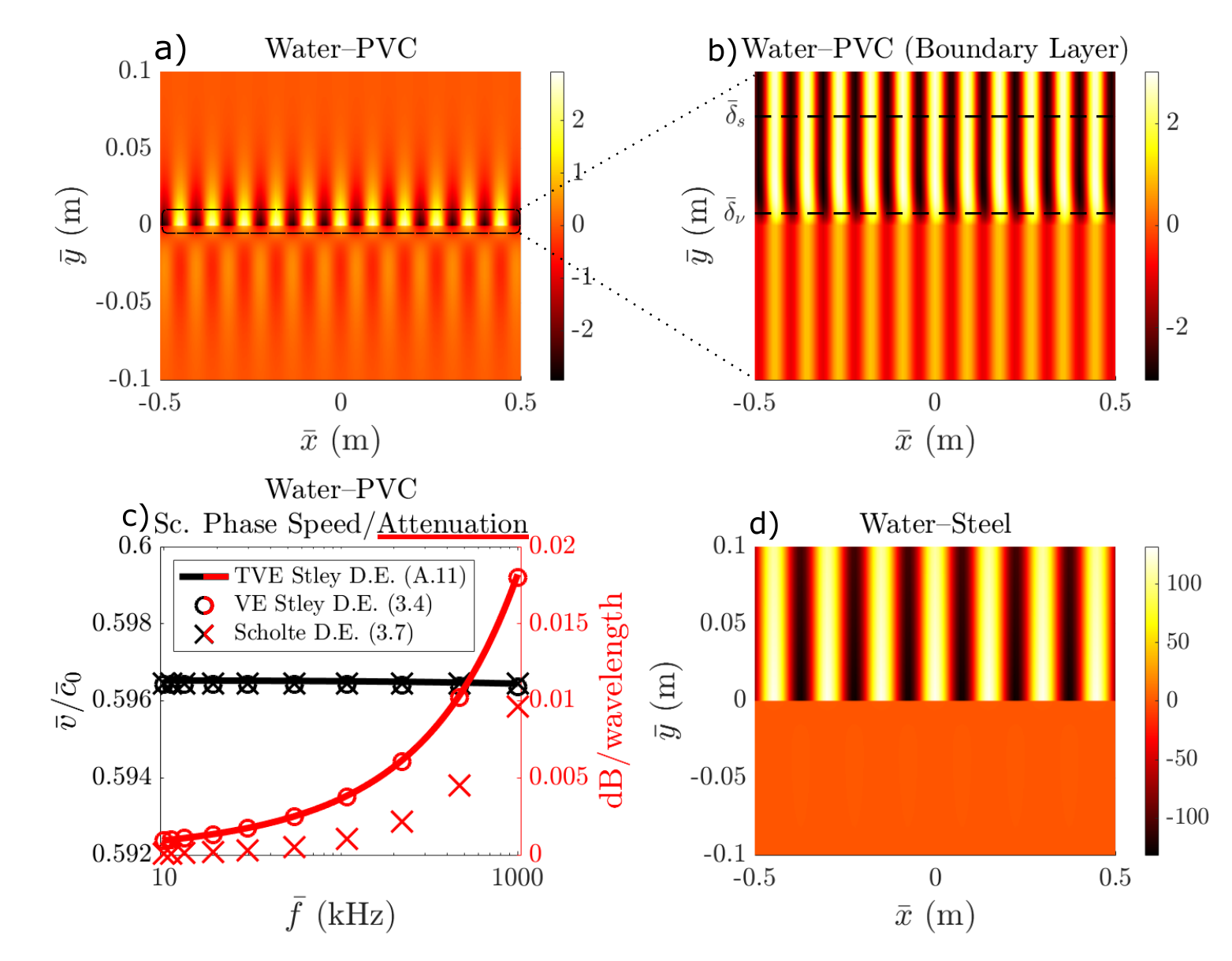}
    \caption{Heatmaps a), b), d) illustrate the dimensionless horizontal particle displacement $\operatorname{Re}\{{u}_x\}$ of the Scholte--Stoneley mode from (\ref{StoneleyDE}) at $10$ kHz for water--PVC (a), (b) and water--Steel (d). (b) corresponds to the same calculation as (a) but near the boundary (as indicated by the small dotted lines) and all three plots have been normalised such that $u_x(y=x=0)=1$. In c) we compare the phase speed and attenuation predicted by the TVE \& VE Stoneley DEs (\ref{stoneley:TVE-TVE}), (\ref{StoneleyDE}), and the Scholte approximation (\ref{ScholteDE}) for the water--PVC case, noting the two different $y$ scales represented by black/red.}
    \label{Fig:Stoneley/Scholte}
\end{figure}

In this case, for the starting point for our DE solver, we choose a real valued $\bar{k}$ whose real part is greater than $\max(\operatorname{Re}\{\bar{k}_\text{F}\}, \operatorname{Re}\{\bar{k}_\Phi\})$. We obtain that the Sc solution to (\ref{StoneleyDE}) for Steel at $10$ kHz gives $\bar{k}_\text{Sc} =42.163 +1.3 \times 10^{-4} \mi$ m$^{-1}$ (noting that $\bar{k}_\text{F} =42.149 +2.5 \times 10^{-6} \mi$ m$^{-1}$), for PVC $\bar{k}_\text{Sc} = 70.70 + 0.0012 \mi$ m$^{-1}$ whereas for $M_{\tau=0.9}$ we obtained $\bar{k}_\text{Sc} = 43.788 + 0.001 \mi$ m$^{-1}$. The difference in the real parts of these roots is remarkable, showing that this PVC is also soft according to the definition used in \cite{staples2021coupled} (following the discussion from Section \ref{subsection:soft/hard solids}). Illustrations of these roots are presented in Figure \ref{Fig:Stoneley/Scholte}. Unlike with the LR above, we can now observe the \textit{trapped} nature of the mode (especially in Fig \ref{Fig:Stoneley/Scholte}a)) propagating parallel to the $y=0$ interface. We nevertheless observe the aforementioned large differences in penetration depths between steel and PVC. With steel Fig \ref{Fig:Stoneley/Scholte}d), the decay length in the water is long and the overall behaviour near the boundary resembles that of a longitudinal wave in the fluid at grazing incidence, whereas for PVC we observe much more motion distributed into the solid, and much shorter decay lengths in the fluid as seen in Fig \ref{Fig:Stoneley/Scholte}a). Given the value of the roots, this can also be easily seen analytically by simply assessing the real part of the various square root functions (\ref{defn: sqrt functions HS}) appearing in (\ref{Potentials:Stoneley}). Similar qualitative results in terms of energies are explicitly given in \cite{favretto1996theoretical} and \cite{staples2021coupled}. From Fig \ref{Fig:Stoneley/Scholte}b) we can observe how the smaller acoustic impedance mismatch between water and PVC also results in a weaker boundary layer effect compared to that for water--steel (as shown for the LR in Fig \ref{Fig:LeakyRayleigh}b)).

As for the LR case, from Fig \ref{Fig:Stoneley/Scholte}c) we again observe that the Scholte DE gives very accurate approximations (with respect to the VE Stoneley DE (\ref{StoneleyDE})) to both the phase speed and attenuation of the Sc mode. { The differences in attenuation in Fig \ref{Fig:Stoneley/Scholte}c) become more noticeable at higher frequencies, with the TVE and VE solutions behaving very similarly.} Note, however the much smaller attenuation values observed for the Sc mode (compared to the LR) since in an ideal fluid--solid interface there is no dissipative mechanism and the root becomes purely real-valued. In Fig \ref{Fig:Stoneley/Scholte}c) there is no notable dispersion in the phase speed of the Sc mode for water--PVC, and we find (not shown) that the same occurs for steel and $M_{\tau=0.9}$ up to MHz frequencies. {Similarly to the LR, including thermal effects predicts slightly higher phase speed values.} Furthermore, so far we have only been considering the KVM such that e.g.\ $\operatorname{Im} \{\bar{\mu}(\bar{\omega}) \} = -\mi \bar{\omega} \bar{\eta}_\mu$ (in both the fluid and solid, for a constant $\bar{\eta}_\mu$) and therefore the larger $\bar{\omega} \bar{\eta}_\mu$, the higher values of dissipation, with the real part remaining constant. For the fluid this is often an accurate model since in this work we are focused on water \cite{nagy1996viscosity}, but this is not generally the case for softer solid media, particularly due to the importance of stress relaxation effects \cite{garcia2022unified}. These are the main topic of discussion in the next section.

\subsubsection{Additional frequency dependence: Influence of stress relaxation} \label{subsection: HS stress relaxation}
As aforementioned, VE effects on the Sc mode for water--synthetic resins were analysed in \cite{favretto1996theoretical}, \cite{favretto1997excitation}. Despite the model used corresponding to a Kelvin--Voigt type model (for fixed frequency), they made acoustical measurements to obtain attenuation coefficients that can be related to viscosity via non-linear functions of frequency. In particular this means that in their case e.g.\ $\bar{\eta}_{\mu} \equiv \bar{\eta}_{\mu}(\bar{\omega})$ (similarly with the bulk viscosity) which becomes a function that must be calculated at each individual frequency and is therefore not constant as in (\ref{eqn: 1D KV}). Furthermore, in their experiments it was found that body wave speeds were almost constant in the range $100$ kHz--$5$ MHz, but they extended this range to cover lower frequencies in their study, so that the sound speeds (and hence the real part of the elastic moduli) are assumed constant in a wide range of frequencies ranging from $20$ kHz--$1$ MHz. Despite these strong assumptions, they obtained accurate predictions of the Sc mode in a number of different scenarios that were confirmed experimentally. In particular, they found low frequency ($10-60$ kHz) dispersion in the water--PVC interface which was not captured with the KVM with the parameters from Table \ref{table: VE parameters}, see Fig \ref{Fig:Stoneley/Scholte}c).\\

In what follows, from a theoretical standpoint we want to include the frequency dependence of the bulk speeds of sound and draw particular attention to the effect of stress relaxation, which is generally most noticeable at lower frequencies \cite{liao2006estimation}. Coincidentally, the importance of these effects were recently pointed out in \cite{staples2021coupled}, who noticed their appearance by experimentally extracting values of the Young's modulus for acrylic using \textit{coupled Scholte modes} (which we discuss in Section \ref{Section: Fluid loaded plates}). Although an analysis of stress relaxation effects is not considered in \cite{staples2021coupled}, it is shown that for a given fluid, the Sc mode is almost independent of the solid's Poisson ratio. More generally, weak frequency dependence of Poisson ratio in VE materials is often found in experiments \cite{hunter1976mechanics}. We will therefore proceed by keeping $\nu_2$ constant and letting the frequency dependence of the Young's modulus $\bar{E}_2(\omega)$ obey the SLSM (\ref{eqn: 1D SLSM}), that is, Material 2 satisfies
 \begin{equation}\label{Young's Mod SLSM}
    {\nu}_2(\omega)=\nu_2, \qquad  \bar{E}_2(\omega) = \bar{E}_\infty - (\bar{E}_0 - \bar{E}_\infty ) \frac{\mi \omega t_r }{1 - \mi \omega t_r},
 \end{equation}
noting that this assumption implies that the solid's shear modulus $\bar{\mu}_2(\omega)$ also satisfies the SLSM with the same relaxation time, and moduli $\bar{\mu}_0, \bar{\mu}_\infty$ (see e.g.\ \cite{obaid2017understanding}). Prony series have shown to be very useful to model the relaxation behaviour of a wide number of VE materials \cite{chen2000determining}. With (\ref{Young's Mod SLSM}) we are employing the simplest case of Prony series by accommodating a single relaxation time $\bar{t}_r$, which in practice is obtained by fitting the model to data from relaxation tests. Nevertheless, as discussed in \cite{garcia2022unified}, it is important to stress that for all real materials, a careful broadband experimental characterization will showcase frequency dependence in all material properties and ought to be taken into account for material-specific studies. Since we do not have easily accessible experimental data, here we will proceed with this assumption and analyze the behaviour resulting from different values of the relevant parameters. We want to observe how the phase speed/attenuation of the modes of propagation can vary according to these parameters and relate the results to those obtained above with local VE. { We ignore thermal coupling in this section, although these can be directly incorporated into the DE (\ref{TVE HS Matrix}) without any additional modelling considerations.}\\

\begin{figure}
    \centering
    \includegraphics[trim={1.5cm 0cm 1.5cm 5cm}, scale=0.45]{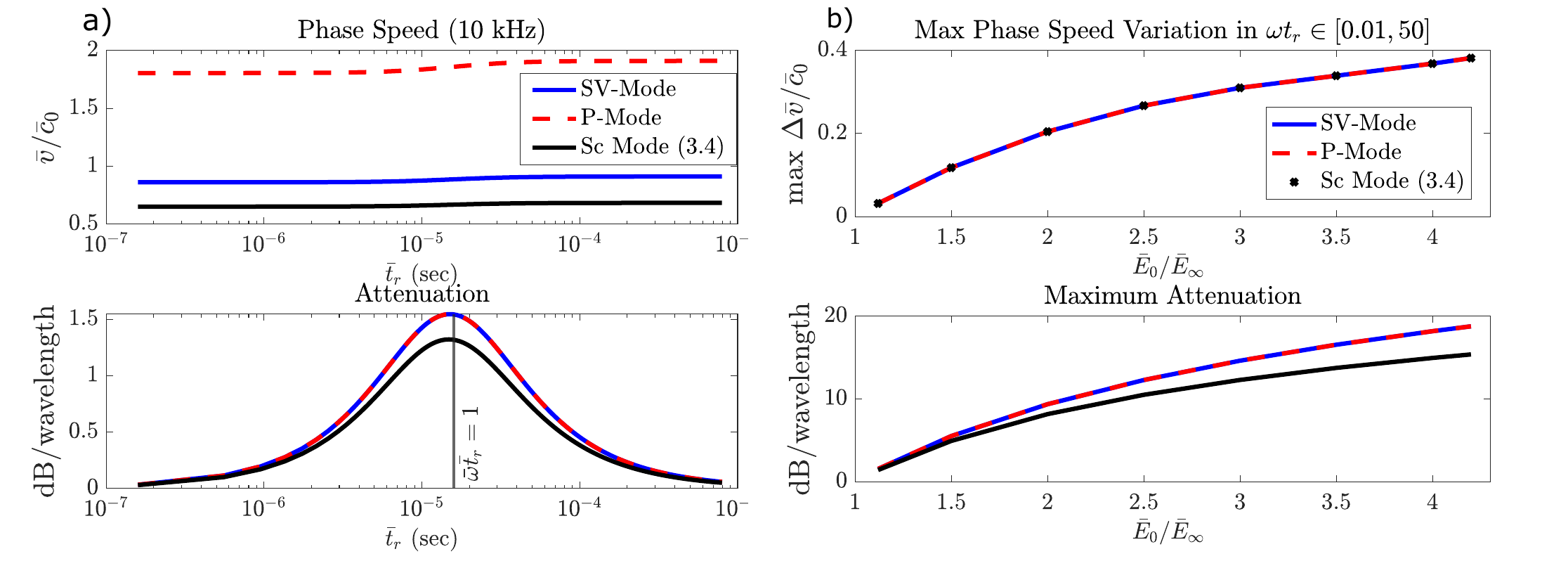}
    \caption{a) Relative phase speed/attenuation of the body and Sc modes following the SLSM at varying relaxation times. b) Maximum phase speed difference and maximum attenuation of the body and Sc modes following the SLSM for increasing values of the ratio $\bar{E}_0/\bar{E}_\infty$. Parameters are specified in (\ref{Parameters for Plot}). Note that the maximum attenuation in a) does not occur exactly at $\omega t_r=1$, since the wavenumber is proportional to the square root of the moduli satisfying the SLSM.}
\label{Fig:Phase/Atten Relaxation HS}
\end{figure}

In Figure \ref{Fig:Phase/Atten Relaxation HS}, we illustrate these effects by plotting the phase speed/attenuation for a particular example with parameters for Material 2
\begin{equation} \label{Parameters for Plot}
     \text{a)} \quad \bar{E}_0/\bar{E}_\infty=1.12, \quad \nu_2=0.3528 \qquad \text{b)} \quad \nu_2=0.3528, \quad {\omega}{t}_r \in [0.01,50] \qquad \text{with } 10 \text{ kHz},
\end{equation}
together with those for water (Material 1) in Table \ref{table: VE parameters}, where we have based the parameters around the PVC sample from Table \ref{table: VE parameters} and assumed that they correspond to the rubbery phase of the material, such that $\bar{E}_\infty = 4.4524$ GPa. In Fig \ref{Fig:Phase/Atten Relaxation HS}a) we observe the phase speed and attenuation as a function of the relaxation time covering the range ${\omega}{t}_r \in [0.01,50]$. We note an expected increase in velocity as the material transitions from rubbery to glassy, as well as a notable global maximum in attenuation close to $\omega t_r =1$ (although not exactly, due to the square root of the elastic moduli appearing in the wavenumbers). In order to obtain the value of this maximum with the KVM used above at 10 kHz, the shear viscosity coefficient of PVC in Table \ref{table: VE parameters} would become as large as $\bar{\eta}_{\mu_2}=1500$ Pa$\cdot$s, illustrating the large differences between the SLSM and KV formulations. In Fig \ref{Fig:Phase/Atten Relaxation HS}b) we provide the maximum phase speed difference (which given our parameters from (\ref{Parameters for Plot}) is given by max $\Delta \bar{v} = \bar{v}_{\omega t_r=50}- \bar{v}_{\omega t_r=0.01}$) as well as the maximum value of attenuation for different ratios $\bar{E}_0/\bar{E}_\infty$. As expected, as this ratio increases the effects observed in Fig \ref{Fig:Phase/Atten Relaxation HS}a) become largely enhanced, noting the larger difference in the attenuation maximum between the body waves and the Sc solution.

These preliminary results illustrate the importance of the Deborah number and equivalently the frequency of operation and approximate relaxation time of the material in question which can cover several orders of magnitude e.g.\ for polyurethane (PU) it can vary between $10-10^3$ (sec) \cite{parnell2019soft}, and therefore these effects would only be noticeable at extremely low frequencies. At higher frequencies (due to the small dispersion of the Sc mode in this regime, e.g.\ Fig \ref{Fig:Stoneley/Scholte}c), \cite{favretto1997excitation}) for fixed material parameters, the behaviour resembles the situation in Fig \ref{Fig:Phase/Atten Relaxation HS}a) with the $x$ axis scaled accordingly. 

\begin{figure}
\centering
\includegraphics[scale=0.45]{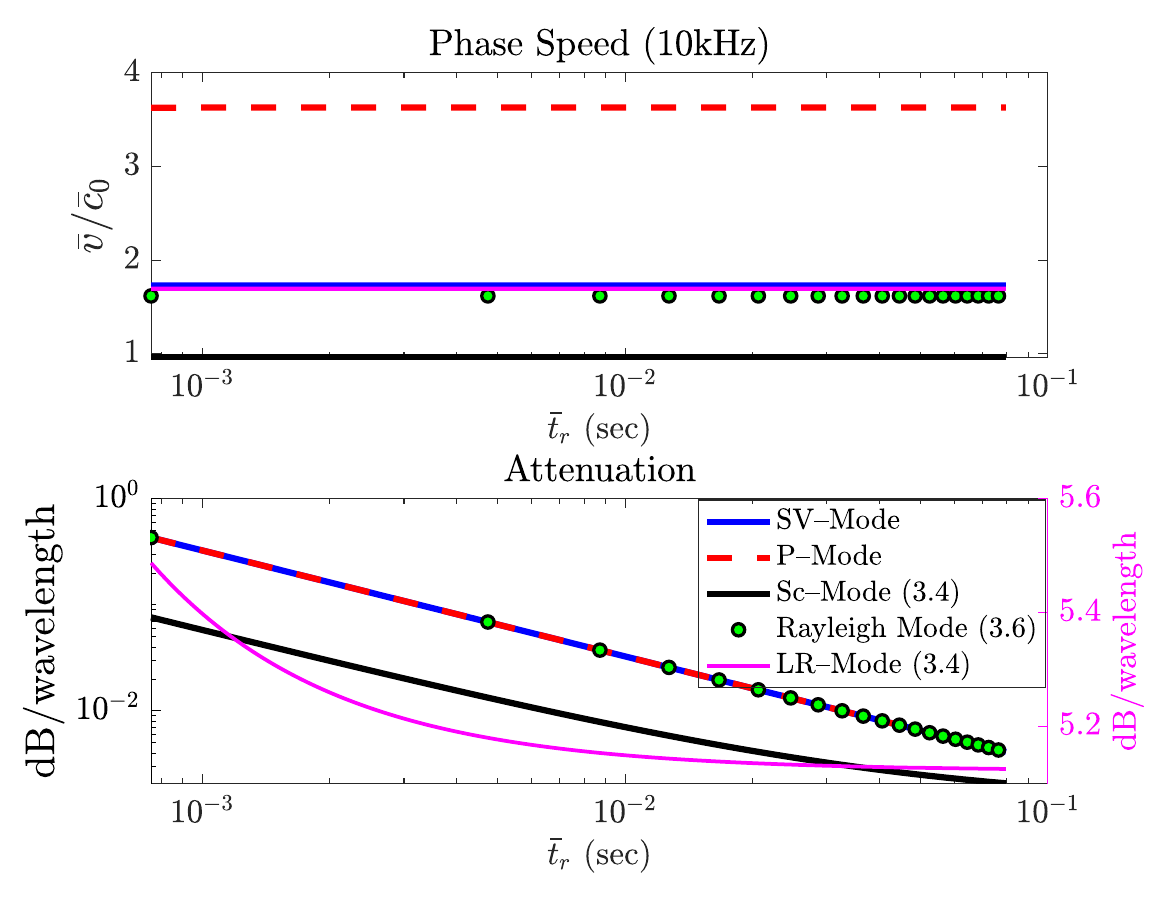}
\caption{Relative phase speed/attenuation of various modes for the ratio $\bar{E}_0/\bar{E}_\infty =4.043$ covering the larger range $\omega t_r \in [50,5000]$ representing the stiffer phase of the material. The magenta $y$ axis in the lower plot corresponds to the magenta curve only (i.e.\ the LR Mode from DE (\ref{StoneleyDE})).}
\label{Fig:Phase/Atten Large Range}
\end{figure}

So far we have only discussed these effects for the Sc mode (and body waves) since, as we saw above the PVC material from Table \ref{table: VE parameters} could not support LR modes when in contact with water. Since this value is now being used as the rubbery phase of the Young's modulus, for a sufficiently large glassy phase ($\bar{E}_0$) the LR should also become a solution. We illustrate this in Figure \ref{Fig:Phase/Atten Large Range} with $\bar{E}_0/\bar{E}_\infty =4.043$, where we give a wider range of relaxation times in the glassy phase of the material, namely  $\omega t_r \in [50,5000]$. The increase in phase speed is therefore almost unnoticeable, nevertheless we can clearly see the decrease in attenuation for the various modes at higher relaxation times. As we saw in Fig \ref{Fig:LeakyRayleigh}c) for $M_{\tau=0.9}$, the LR mode has a much higher attenuation due to energy radiation, so a separate scale is given in Fig \ref{Fig:Phase/Atten Large Range}  (on the right, in cyan) to represent its decrease. From a practical perspective, this also shows that stress relaxation effects can be noticed far away from glass transition, especially the larger the ratio $\bar{E}_0/\bar{E}_\infty$ becomes.\\

It is worth remarking that the dispersive behaviour illustrated in Figures \ref{Fig:Phase/Atten Relaxation HS}, \ref{Fig:Phase/Atten Large Range} and discussed above requires a viscoelastic model that captures stress relaxation such as the SLSM employed here, as opposed to e.g.\ the standard KVM discussed earlier. Naturally, by using measurements to develop a frequency dissipation coefficient $ \bar{\eta}_{\mu}(\bar{\omega})$ as employed in \cite{favretto1997excitation} (rather than the constant $\bar{\eta}_{\mu}$ in the standard KVM), the attenuative properties can be accurately predicted, however it cannot capture the changes in phase speed observed in the top figures of Fig \ref{Fig:Phase/Atten Relaxation HS}a),b). Therefore, if the frequencies are sufficiently high this becomes a good approximation as seen from the top of Figure \ref{Fig:Phase/Atten Large Range}, but care must be taken in the lower frequency range, as discussed in e.g.\ \cite{staples2021coupled} and \cite{liao2006estimation}.


\newpage 

\section{Double interface: VE waveguides/plates bounded by semi-infinite VE continua}\label{Section: Double interface general}

Having analysed the acoustic surface mode solutions to the initial single interface configuration consisting of two semi-infinite half-spaces in 2D, we next consider an additional parallel interface separated from the first interface by a distance of $\bar{W} = 2\bar{L}$, as shown in the right of Figure \ref{Fig:Geometries}. Our framework will conveniently allow us to study the dissipative mechanisms in two physically different problems under the same set of dispersion equations, namely fluid-filled channels within solids and fluid-loaded plates; analysed in Sections \ref{Section: Fluid channels}, \ref{Section: Fluid loaded plates} respectively. The relevant DEs are discussed shortly, but it will be convenient to first non-dimensionalise the problem.

\subsection{Non-dimensionalisation}\label{subsection:non-dimensionalization}
Unlike in the case of a single interface, in the current geometry there is a clear length scale dictated by the finite dimension of Material 1. It is therefore convenient to non-dimensionalise the relevant equations, which we do via
\begin{align*}
    & \Delta = \bar{L}^2 \bar{\Delta} \quad \omega = \frac{\Bar{L} \Bar{\omega}}{\Bar{c}_\square}, \quad  \{\mathbf{u}_m,\mathbf{x}\}=\frac{1}{\bar{L}} \{\bar{\mathbf{u}}_m,\bar{\mathbf{x}}\},\quad \{\bm{\sigma}_m,\lambda_m, \mu_m, E_m,\mathcal{Q}\}=\frac{1}{\bar{\rho}_1 \bar{c}^2_\square} \{\bar{\bm{\sigma}}_m,\bar{\lambda}_m, \bar{\mu}_m, \bar{E}_m, \bar{\mathcal{Q}}\}, \\
    & \rho_s = \frac{\bar{\rho}_2}{\bar{\rho}_1}, \quad \{k,k_{\phi_m}, k_{\Phi_m},\gamma_{\phi_m}, \gamma_{\Phi_m}\} = \bar{L} \{\bar{k},\bar{k}_{\phi_m}, \bar{k}_{\Phi_m}, \bar{\gamma}_{\Phi_m},\bar{\gamma}_{\phi_m}\}, \quad  \{c,c_{\phi_m},c_{\Phi_m}\}=\frac{1}{\bar{c}_\square}\{\bar{c},\bar{c}_{\phi_m},\bar{c}_{\Phi_m}\}, \\
    &  t_r =\bar{c}_\square \bar{t}_r/\bar{L}, \quad \{\phi_m,\Phi_m\} = \frac{1}{\bar{L}^2}\{\bar{\phi}_m,\bar{\Phi}_m\}, \quad  \{\eta_{\mu_m},\eta_{K_m}\} = \frac{1}{\bar{\rho}_1 \bar{c}_\square \bar{L}}\{\bar{\eta}_{\mu_m},\bar{\eta}_{K_m}\},
\end{align*}
where $m=1,2$. The (constant) sound speed ($\bar{c}_\square$) will be chosen accordingly in Sections \ref{Section: Fluid channels}, \ref{Section: Fluid loaded plates}.

\subsection{The generalised dispersion equations}
We set up the 2D coordinate system such that the $x$ direction is aligned with the interfaces and $y=0$ lies in the middle of Medium 1, and we therefore must ensure that the continuity of traction and displacement boundary conditions (\ref{VE 2 HS BCs 1}) are satisfied at ${y} = \pm {1}$. By exploiting the symmetry of the configuration about ${y}=0$, the general $8 \times 8$ system arising from the continuity BCs on each interface can be split into two independent $4 \times 4$ systems (see e.g.\ \cite{osborne1945transmission}). Indeed, \textit{symmetric modes} require ${u}_x^1$, $\phi_1$ to be even functions of ${y}$ whereas ${u}_y^1$, $\Phi_1$ must be odd functions of ${y}$, therefore the potentials in (\ref{viscoelastic helmholtz decomposition}) are given by

\begin{subequations} \label{Symm. Waveguides}
\begin{align} \label{Symm. Waveguide Compressional}
\phi_1 &= {\text{P}1}_\text{S} \cosh{(\gamma_{\phi_1} y)} \me^{\mi k x},  &
 \phi_2(x,y) = 
\begin{cases}
		  {\text{P}2}_\text{S}\me^{  \gamma_{\phi_2} (y+1) + \mi k x}, \qquad & y \leq -1\\
          {\text{P}2}_\text{S}\me^{  -\gamma_{\phi_2} (y-1) + \mi k x}, \qquad & y \geq  1
\end{cases},\\ \label{Symm. Waveguide Shear}
\Phi_1 &= {\text{S}1}_\text{S} \sinh{(\gamma_{\Phi_1} y)} \me^{\mi k x},  &
\Phi_2(x,y) = 
\begin{cases}
		  {\text{S}2}_\text{S}\me^{  \gamma_{\Phi_2} (y+1) + \mi k x}, \qquad & y \leq -1\\
          {\text{S}2}_\text{S}\me^{  -\gamma_{\Phi_2} (y-1) + \mi k x}, \qquad & y \geq  1
\end{cases},
\end{align}
\end{subequations}
for some complex valued amplitudes ${\text{P}1}_\text{S}, {\text{P}2}_\text{S}, {\text{S}1}_\text{S},  {\text{S}2}_\text{S}$. Given (\ref{Symm. Waveguides}), it can be shown that symmetric modes are given by solutions to
\begin{equation}\label{symmetricVE channel :VE-VE-VE}
    \begin{aligned}
        & c^4\left( ({1}-{\rho_s})^2 \tanh \left(\gamma_{\Phi_1} \right) -{A_1} {\rho_s} \tanh \left(\gamma_{\phi_1} \right)
   \left({B_2} \tanh \left(\gamma_{\Phi_1} \right)+{B_1}
   {\rho_s}\right)-{A_2} {B_2}  \tanh \left(\gamma_{\Phi_1} \right)-{A_2} {B_1} {\rho_s}\right) \\
   &+2 c^2 {\mathcal{Q}} \left(-{\rho_s} {A_1} {B_1}  
   \tanh \left(\gamma_{\phi_1} \right)+( {A_2} {B_2} - 1+{\rho_s}) \tanh \left(\gamma_{\Phi_1} \right)\right)\\
   &+{\mathcal{Q}}^2 \left({A_1} {B_1} \tanh \left(\gamma_{\phi_1} \right)-\tanh \left(\gamma_{\Phi_1} \right)\right)  ({A_2}
   {B_2}-1)= 0,
    \end{aligned}
\end{equation}
which was derived in Mathematica, recalling that the quantities $A_1,A_2,B_1,B_2, \mathcal{Q}$ are defined in (\ref{definition: HS quantities}) (and noting the non-dimensionalisation above). Conversely, \textit{anti-symmetric modes} require ${u}_x^1$, $\phi_1$ to be odd functions of ${y}$ and ${u}_y^1$, $\Phi_1$ to be even, such that
\begin{subequations}\label{antiSymm. Waveguide}
\begin{align} \label{antiSymm. Waveguide Compressional}
\phi_1 &= \text{P}1_\text{A} \sinh{(\gamma_{\phi_1} y)} \me^{\mi k x},  
& \phi_2(x,y) = 
\begin{cases}
		 - {\text{P}2}_\text{A}\me^{  \gamma_{\phi_2} (y+1) + \mi k x}, \qquad & y \leq -1\\
          {\text{P}2}_\text{A}\me^{  -\gamma_{\phi_2} (y-1) + \mi k x}, \qquad & y \geq  1
\end{cases},\\
\label{antiSymm. Waveguide Shear}
\Phi_1 &= \text{S}1_\text{A} \cosh{(\gamma_{\Phi_1} y)} \me^{\mi k x}, & \Phi_2(x,y) = 
\begin{cases}
		  {\text{S}2}_\text{A}\me^{  \gamma_{\Phi_2} (y+1) + \mi k x}, \qquad & y \leq -1\\
          {\text{S}2}_\text{A}\me^{  -\gamma_{\Phi_2} (y-1) + \mi k x}, \qquad & y \geq  1
\end{cases},
\end{align}
\end{subequations}
for complex valued amplitudes ${\text{P}1}_\text{A}, {\text{P}2}_\text{A}, {\text{S}1}_\text{A},  {\text{S}2}_\text{A}$. Similarly, given (\ref{antiSymm. Waveguide}) it can be shown that anti-symmetric modes are given by solutions to
\begin{equation}\label{ANTIsymmetricVE channel :VE-VE-VE}
    \begin{aligned}
& c^4 \left( (1-{\rho_s})^2\tanh \left(\gamma_{\phi_1} \right) - B_1 \rho_s \tanh  \left(\gamma_{\Phi_1} \right) \left({A_2} \tanh \left(\gamma_{\phi_1} \right) +  {A_1} {\rho_s} \right) - {A_2} {B_2}
    \tanh \left(\gamma_{\phi_1} \right) - A_1{B_2}
 {\rho_s} \right) \\ &+ 2 c^2 {\mathcal{Q}} \left( - {\rho_s} {A_1}
   {B_1}  \tanh \left(\gamma_{\Phi_1} \right) + ({A_2}
   {B_2}- 1+{\rho_s}) \tanh \left(\gamma_{\phi_1} \right)\right)\\  &+ {\mathcal{Q}}^2
    \left({A_1}
   {B_1} \tanh \left(\gamma_{\Phi_1} \right) - \tanh \left(\gamma_{\phi_1} \right)\right)({A_2} {B_2}-1) = 0,
\end{aligned}
\end{equation}
and it is straightforward to check that in the short-wavelength limit $\gamma_{\phi_1} $, $\gamma_{\Phi_1}  \gg 1$ both (\ref{symmetricVE channel :VE-VE-VE}), (\ref{ANTIsymmetricVE channel :VE-VE-VE}) recover the Stoneley DE (\ref{StoneleyDE}) (since in this limit there is no distinction between symmetric and anti-symmetric motion). This natural geometric consequence will allow us to use the knowledge of the two semi-infinite half-space configuration discussed in Section \ref{Section: Single Interface} in order to obtain the initial behaviour within the slit/plate. { We show in Appendix \ref{appendix:section: TVE Slit/Plate} that when thermal effects are present, (\ref{symmetricVE channel :VE-VE-VE}), (\ref{ANTIsymmetricVE channel :VE-VE-VE}) take the more general forms (\ref{TVE-TVE-SYM}), (\ref{TVE-TVE-ANTISYM}) which require taking the determinant of $6 \times 6$ matrices.} Having obtained the general dispersion equations for natural modes of interest (\ref{symmetricVE channel :VE-VE-VE}), (\ref{ANTIsymmetricVE channel :VE-VE-VE}), we will next focus on the analysis of the two distinct limits of physical interest here.

\subsection{Limit A): Fluid-filled channels within semi-infinite solids} \label{Section: Fluid channels}
As aforementioned, the effects of visco-thermal boundary layers and FSI in narrow water/air-filled slits were examined extensively in \cite{cotterill2018thermo}. In the FSI analysis therein, the only solid material considered was steel, which given the discussion above constitutes a hard solid when in contact with water (for air, in all cases FSI effects were negligible). In this section, we explore the differences arising when soft TVE media are instead considered.

Following \cite{cotterill2018thermo}, the analysis below is focused only on the lowest order symmetric mode, so that here we will only be considering the roots of (\ref{symmetricVE channel :VE-VE-VE}). The rationale employed in \cite{cotterill2018thermo} was that this was sensible since for an inviscid fluid with rigid BCs it is the only propagating mode at our region of interest; namely thin channel widths/low frequencies. Of course, FSI effects arising from the consideration of solid media in contact with water are such that the rigid BC does not apply, and the current situation is far from this idealized scenario, since we are also analysing softer solid media. Nevertheless, the main objective of this section is to extend the results presented in \cite{cotterill2018thermo} by including the effects of TVE losses within the (infinite) solid bounding the fluid, together with the previously considered boundary layer effects.

As we will see shortly, the mode in consideration is in fact strongly related to the Sc mode from Section \ref{subsubsec: Sch Mode}, which becomes coupled in the slit region as the thickness decreases (with respect to the transverse wavelength). As a result, we are essentially studying the `coupled duct--Scholte mode' including dissipation. On the other hand, unlike for the half-space, geometrical confinement implies that LR waves cannot radiate energy away from the interfaces and hence propagate plane-wave like with little attenuation along the channel. Therefore, as opposed to the coupled duct--Scholte mode, they are only expected to propagate in sufficiently wide channels \cite{cotterill2018thermo}.\\

Given the non-dimensionalisation in Section \ref{subsection:non-dimensionalization}, in this Section we let $\bar{c}_\square \equiv \bar{c}_0 = 1490$ m/s i.e.\ we choose the adiabatic speed of sound in water at 10 degrees Celsius and $\bar{\rho}_1=1000$ kg/m$^3$ (following Table \ref{table: VE parameters}). With regards to the choice of VE frequency dependence, we will follow a similar structure to Section \ref{Section: Single Interface} for the single interface: first we assume the KVM for the solid according to Table \ref{table: VE parameters}, and then we consider the SLSM in  Section \ref{subsection: Slit stress relaxation}.

\subsubsection{Phase speed and attenuation}\label{subsection: slit phase speed / attenuation}
\begin{figure}
    \centering
    \includegraphics[trim={0cm 0cm 0cm 0cm}, scale=0.41]{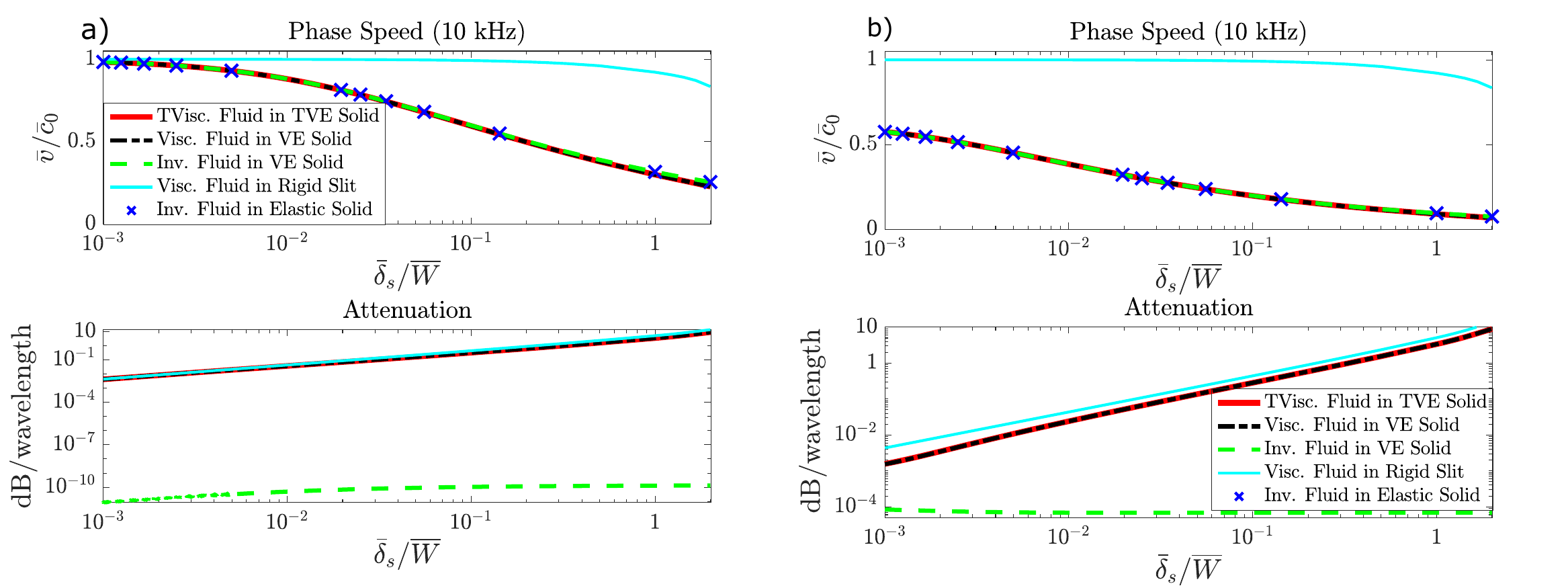}
    \caption{Phase Speed/Attenuation at $10$ kHz for water-filled channels of decreasing width within semi-infinite a) Steel, and b) PVC media, according to the KVM with material properties in Table \ref{table: VE parameters} (and Table \ref{table: Th parameters} for the red curves). In particular, we note the large difference in the initial value of the phase speed between a) and b) which is dictated by the Stoneley DE (\ref{StoneleyDE}).}
\label{Fig:Phase/Atten Fluid filled channel}
\end{figure}

\begin{figure}
    \centering
    \includegraphics[trim={0cm 0cm 0cm 0cm}, scale=0.41]{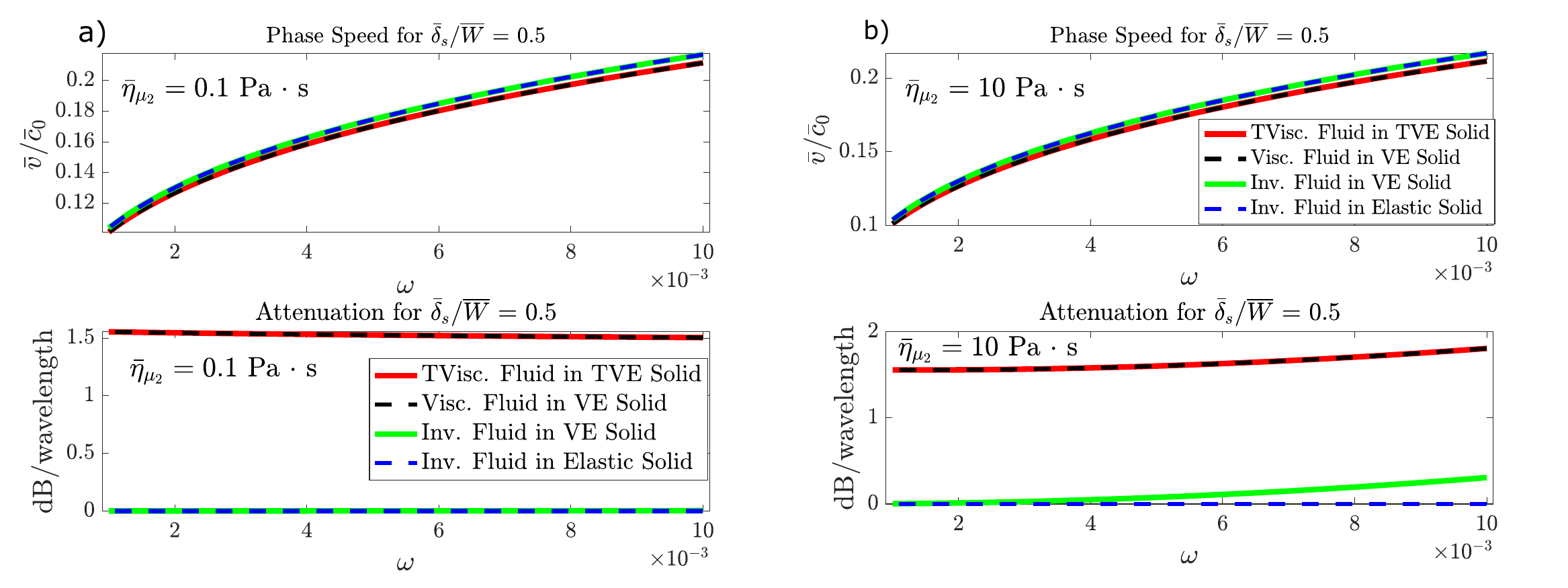}
    \caption{Relative phase speed/attenuation at fixed channel width $\bar{\delta}_s/\bar{W}=0.5$ for increasing frequency in a water-filled PVC channel following the KVM for: a) Shear PVC viscosity coefficient $\eta_{\mu_2}$ from Table \ref{table: VE parameters}, b) Larger shear PVC viscosity coefficient. In a) we observe dispersion in phase speed, but not in attenuation, whereas in b) we observe an additional VE frequency dependence in attenuation.}
\label{Fig:Phase/Atten water filled PVC channel varying omega}
\end{figure}

In order to find the roots of (\ref{symmetricVE channel :VE-VE-VE}) using \textit{fsolve}, we make use of an iterative numerical scheme, whose starting point is dictated by the solutions to the Stoneley DE (\ref{StoneleyDE}), corresponding to the wide channel limit, as noted just below (\ref{ANTIsymmetricVE channel :VE-VE-VE}) {(similarly for the case when thermal coupling is considered; to find the roots of (\ref{TVE-TVE-SYM}), the starting point is dictated by the solutions to (\ref{stoneley:TVE-TVE}))}. We subsequently gradually reduce the channel widths up to the values of interest, by using the root found for the prior larger value of $\bar{W}$ as a new starting point for \textit{fsolve}. In Figure \ref{Fig:Phase/Atten Fluid filled channel} we observe the phase speed/attenuation as a function of channel width in terms of $\bar{\delta}_s/\bar{W}$ at $10$ kHz for a) water--steel and b) water--PVC. We directly observe the difference in phase speed between the hard/soft solids at wide channel widths. In particular, the mode in water--PVC is highly subsonic, as we observed in Fig \ref{Fig:Stoneley/Scholte}c) for the half-space Sc mode. More generally, it is clear that FSI effects in both materials dramatically reduce the phase speed along the channel as opposed to the rigid case. { We observe that for the parameters used in Figs \ref{Fig:Phase/Atten Fluid filled channel} a),b), phase speed and attenuation among the thermo-viscous fluid--TVE interface (red curves) remains almost identical to the viscous fluid--VE results (black dashed curves). The rigid BC results (cyan curve) where the solid motion is ignored is plotted for reference.} For further comparisons we give additional curves that neglect particular physical effects. In particular, for the inviscid fluid we assume medium 1 is s.t. $\mu_1(\omega) \rightarrow 0$, implying that $k_{\Phi_1}^2 \rightarrow \infty$ and hence $\gamma_{\Phi_1}, B_1 \rightarrow \infty$ whereas $\mathcal{Q}\rightarrow -2  \rho_s c_{\Phi_2}^2$ and (\ref{symmetricVE channel :VE-VE-VE}) simplifies to
\begin{align}
      \label{inv-VE 72}
      \tanh(\gamma_{\phi_1}) \left[(2k^2 - k_{\Phi_2}^2)^2 - 4 k^2 \gamma_{\phi_2} \gamma_{\Phi_2} \right] + \frac{\gamma_{\phi_2} k_{\Phi_2}^4}{\gamma_{\phi_1}  \rho_s}=0, 
\end{align}
noting the similarity with the limit taken to arrive at the Scholte DE (\ref{ScholteDE}) from the Stoneley DE (\ref{StoneleyDE}) in the half-space configuration. Equation (\ref{inv-VE 72}) is identical to equation (72) in \cite{cotterill2018thermo} except here in the parameters for medium 2 we are including frequency dependence (barring the blue crosses in Figures \ref{Fig:Phase/Atten Fluid filled channel}, \ref{Fig:Phase/Atten water filled PVC channel varying omega} where we further let $\eta_{\mu_2}, \eta_{K_2} \equiv 0$ in the solid which yields real valued roots). {Nevertheless, the VE losses in both solids remain small, as can be seen from the green curves in the bottom plots from Fig \ref{Fig:Phase/Atten Fluid filled channel}, which shows that the great majority of the attenuation is indeed due to the boundary layers in water (red/black curves).} Note that this is expected for steel at these frequencies due to its negligible viscosity, but less so for PVC (Table \ref{table: VE parameters}). When performing similar calculations at higher frequencies, we find that the phase speed values for most (fixed) channel widths become larger, whilst the attenuation remains fairly constant. This is shown explicitly for water--PVC in Figure \ref{Fig:Phase/Atten water filled PVC channel varying omega}a) for the particular case when $\bar{\delta}_s/\bar{W}=0.5$, noting that a similar frequency dependence is obtained for steel in \cite{cotterill2018thermo}. Dispersion in attenuation can nevertheless be seen when a higher coefficient of viscosity is employed, as illustrated in Figure \ref{Fig:Phase/Atten water filled PVC channel varying omega}b) where we used $\bar{\eta}_{\mu_2}=10$ Pa$\cdot$s, although it has close to no effect on the relative phase speed. Finally, following our initial discussion above, we can only observe the coupled LR--duct mode for water--steel at $10$ kHz up to $\bar{\delta}_s/\bar{W} \approx 0.03$; for thinner channels it becomes cut-off (not shown).

\subsubsection{Displacement fields}
\begin{figure}
\centering
   \includegraphics[scale=0.45]{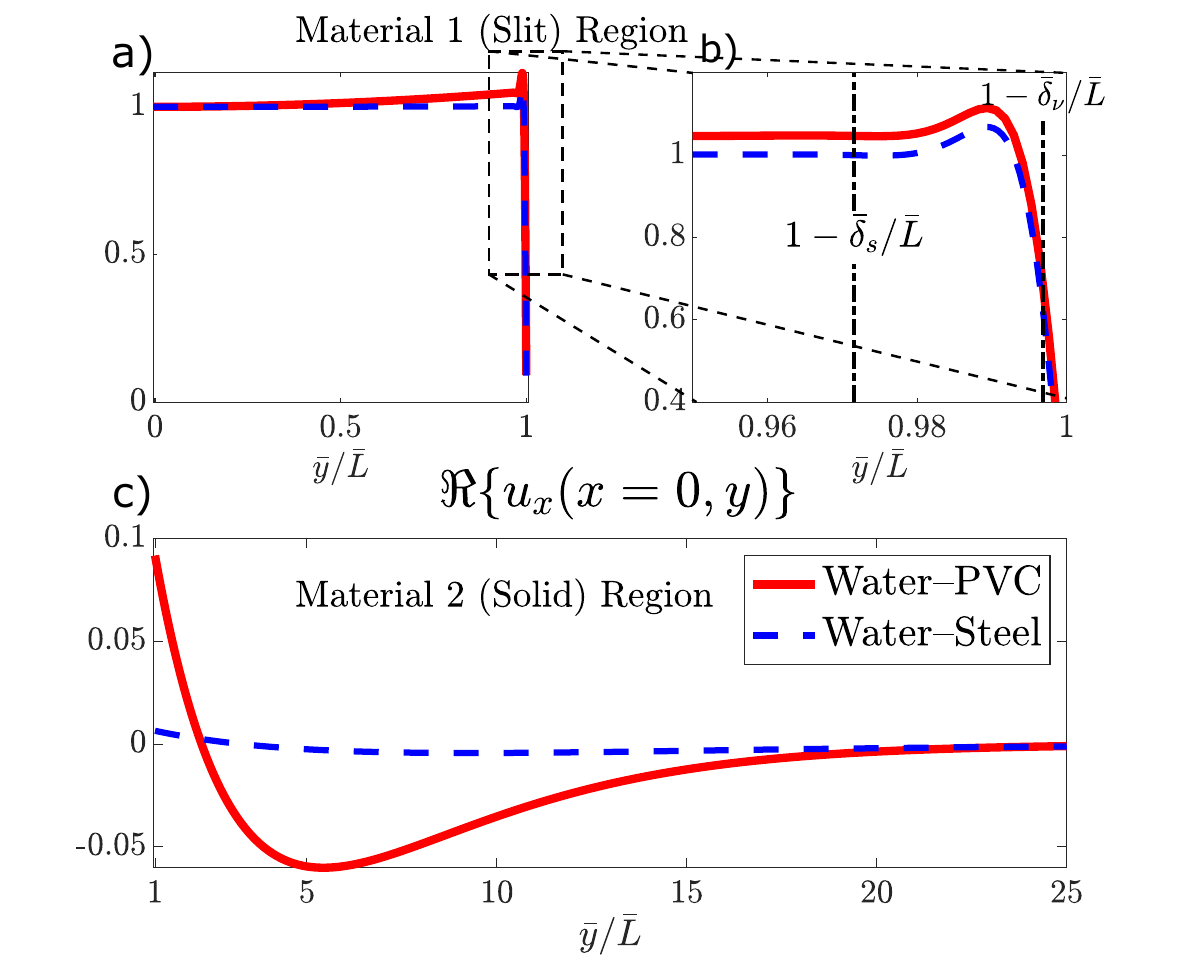}
\caption{Real part of the horizontal displacement fields $\operatorname{Re}\{u_x(x=0,y)\}$ for the lowest order symmetric mode propagating in a viscous water-filled channel at $100$ kHz within steel/PVC, for a channel width of $\bar{W} = 70 \bar{\delta}_s$. a) represents the fluid region $0 \leq \bar{y} \leq \bar{L}$, b) the fluid behaviour near the boundary, and c) the solid region $ \bar{L} \leq \bar{y} \leq 25 \bar{L}$. In each case the plots have been normalized such that  $u_x(y=x=0)=1$.}
\label{Fig:Displacements fluid filled channels}
\end{figure}

The particle motion of the mode under consideration for a water--steel interface was studied in detail in \cite{cotterill2018thermo}. Here we simply want to illustrate whether any significant changes occur when steel is replaced by a softer medium. In Figure \ref{Fig:Displacements fluid filled channels} we give direct comparisons between steel/PVC of the horizontal displacement $\operatorname{Re}\{u_x\}$ for a wide channel, represented by $\bar{W} = 70 \bar{\delta}_s$. The behaviour in the fluid region, Fig \ref{Fig:Displacements fluid filled channels}a),b) is rather similar between the two interfaces, noting that the boundary layer region is well approximated by  Stokes' boundary layer $\bar{\delta}_s/\bar{L}$. Nevertheless, in Fig \ref{Fig:Displacements fluid filled channels}a) we also observe a slight decay from the boundary region towards the centre of the channel for PVC which is not observed in steel. We observe (not shown) that this feature becomes increasingly noticeable at wider channel widths. This is expected, since as the channel width becomes larger, the solution begins to resemble that of the half-space Sc mode localized in each boundary, and as we observed above, the decay length in the fluid is much longer for hard solids (Fig \ref{Fig:Stoneley/Scholte}d)) than it is for softer media (Fig \ref{Fig:Stoneley/Scholte}a)). On the other hand, in the solid region  Fig \ref{Fig:Displacements fluid filled channels}c), we see a large difference between the motion in PVC and steel, with the displacement of the former being more than an order of magnitude larger than the latter at the interface $y=\pm 1$. Nevertheless, this motion still decays rapidly within the solid, which is expected since we observed in Figure \ref{Fig:Phase/Atten Fluid filled channel}b) that there was no damping due to radiation loss in the form of elastic waves through the interfaces into the solid. At narrower channel widths (not shown) the motion at the interface with the solid region becomes increasingly reduced, although the difference between hard/solid can still be appreciated.

\subsubsection{Additional frequency dependence: Influence of stress relaxation}\label{subsection: Slit stress relaxation}
As was done for the half-space, we now wish to assess whether the inclusion of stress relaxation effects via the SLSM in the \textit{host} solid medium can alter the phase speed/attenuation of the Sc--duct mode under consideration. We observed in Figure \ref{Fig:Phase/Atten water filled PVC channel varying omega} how, when employing the KVM with sufficiently high viscosity coefficient, the attenuation increases (unboundedly) with $\omega$, but this has no effect on the corresponding relative phase speed. We will follow the assumptions from Section \ref{subsection: HS stress relaxation} and assume a constant Poisson's ratio and let the Young's modulus (of Medium 2) obey the SLSM as in (\ref{Young's Mod SLSM}), noting that under the current non-dimensionalization the relaxation time is given by $ t_r =\bar{c}_0 \bar{t}_r/\bar{L}$. {As for the half-space, we do not discuss thermal coupling in this section.}

\begin{figure}
\centering
   \includegraphics[scale=0.38]{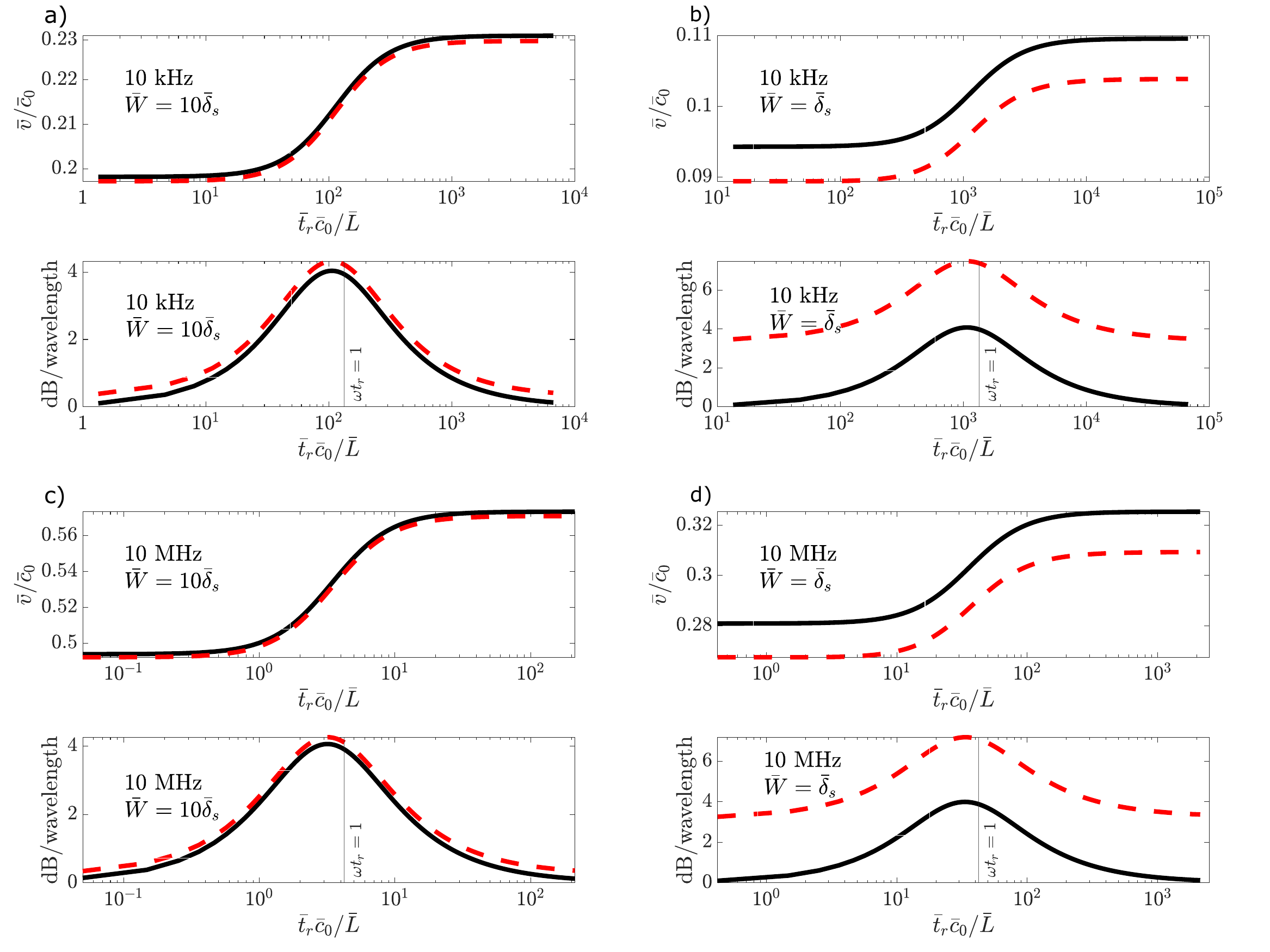}
\caption{Relative phase speed/attenuation in a water-filled PVC slit obeying the SLSM, as a function of non-dimensional relaxation time $t_r$ for various channel widths and frequencies. The red-dashed curve represents viscous water given by (\ref{symmetricVE channel :VE-VE-VE}), whereas the black curve is for inviscid water and is therefore a solution to (\ref{inv-VE 72}). In all cases we have $\bar{E}_0/\bar{E}_\infty=1.572$, and the relaxation times chosen cover the range $\omega t_r \in [0.01,50]$.}
\label{Fig:SLIT RELAXATION 4 COMBINED}
\end{figure}

For the illustrations, again we assume that the rubbery phase of the PVC material corresponds to the value given in Table \ref{table: VE parameters}, so that $\bar{E}_\infty=4.4524$ GPa. In Figure \ref{Fig:SLIT RELAXATION 4 COMBINED} we give the phase speed/attenuation at various frequencies and channel widths as a function of relaxation time $t_r$, chosen in each case so that the range $\omega t_r \in [0.01,50]$ is covered, and therefore the glass transition of the material is showcased. The qualitative behaviour is therefore very similar within Figures \ref{Fig:SLIT RELAXATION 4 COMBINED}a)--d), where we note the increasing phase speeds as the material becomes glassier, and the global maximum attenuation near $\omega t_r=1$, as we saw in Figure \ref{Fig:Phase/Atten Relaxation HS}a) for the half-space. Nevertheless, in this case there are significant changes in the relevant values when the main parameters vary, following what we saw in Figures \ref{Fig:Phase/Atten Fluid filled channel}, \ref{Fig:Phase/Atten water filled PVC channel varying omega}. That is, with higher frequencies (smaller channel widths) we observe overall higher values in phase speed, and the effect of the fluid's viscosity becomes particularly relevant at narrower channel widths. Furthermore, as we observed for the half-space in Figure \ref{Fig:Phase/Atten Relaxation HS}b), the larger the ratio $\bar{E}_0/\bar{E}_\infty$ gets, the larger the maximum phase speed variation and maximum attenuation becomes. 

\begin{figure}
\centering
\includegraphics[scale=0.40]{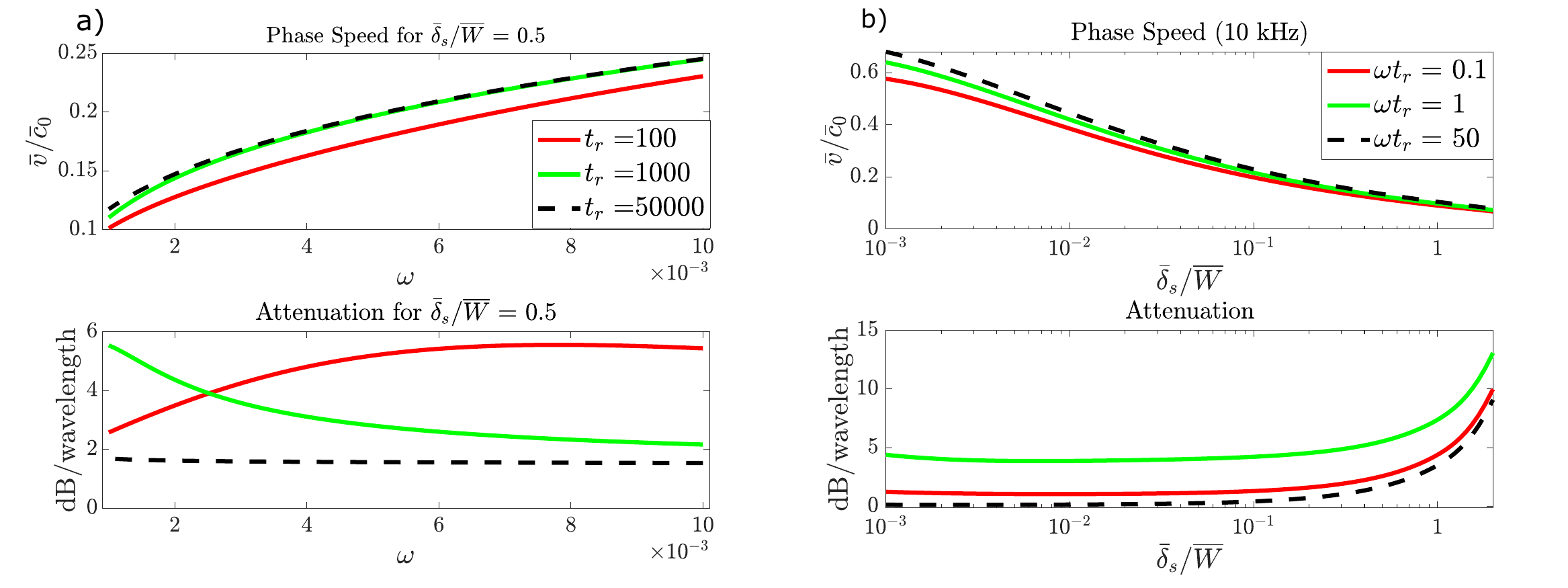}
\caption{Relative phase speed/attenuation in a (viscous) water-filled PVC slit obeying the SLSM, with $\bar{E}_0/\bar{E}_\infty=1.572$: a) as a function of frequency for a fixed channel width of $\bar{\delta}_s/\bar{W}=0.5$, b) as a function of channel width with dimensional frequency $10$ kHz. In each case the relaxation times have been chosen to represent the differences between the glassy/rubbery phases of the solid medium, noting the differences in attenuation in a) for the smaller relaxation times as the frequency increases.}
\label{Fig:SLIT RELAXATION PLOT 2}
\end{figure}

\begin{figure}
\centering
   \includegraphics[scale=0.40]{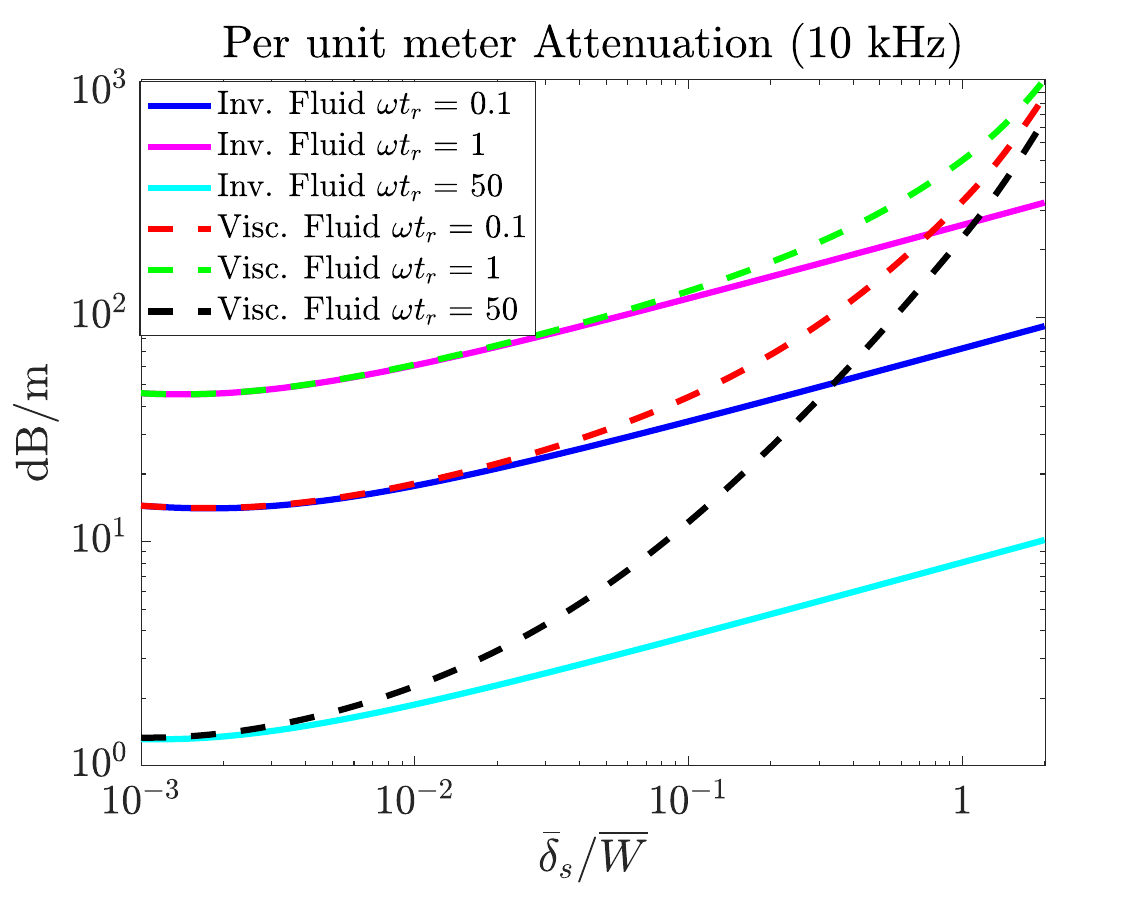}
\caption{Attenuation per unit metre in a water-filled PVC slit obeying the SLSM, with $\bar{E}_0/\bar{E}_\infty=1.572$. In each case the relaxation times have been chosen to represent the differences between the glassy/rubbery phases of the solid medium. Viscous (Visc.) fluid solutions are obtained from (\ref{symmetricVE channel :VE-VE-VE}), whereas the inviscid (Inv.) fluid solutions are from (\ref{inv-VE 72}).}
\label{Fig:per unit metre attenuation}
\end{figure}

In Figure \ref{Fig:SLIT RELAXATION PLOT 2} we give similar plots to those in Figs \ref{Fig:Phase/Atten Fluid filled channel}, \ref{Fig:Phase/Atten water filled PVC channel varying omega} for various fixed relaxation times. In Figure \ref{Fig:SLIT RELAXATION PLOT 2}a) the chosen $t_r$ are such that $\omega t_r = {0.1,1,50}$ at the initial value of frequency, so that at this (initial) frequency we are covering the rubbery/glassy phases and glass transition. As a result, we observe how as frequency increases the attenuation of the initially rubbery phase ($t_r=100$) undergoes glass transition, whereas naturally in the two other cases ($t_r=1000, 50000$) an increase of frequency results in a reduction of attenuation (per wavelength). From Figure \ref{Fig:SLIT RELAXATION PLOT 2}b) we observe how the initial phase speed is higher in the glassy phase, and particularly how with $\omega t_r$ fixed, VE relaxation effects together with boundary layers can lead to a significant increase in attenuation (compared to e.g.\ Fig \ref{Fig:Phase/Atten Fluid filled channel}b)). Note that the small decrease in attenuation observed for wide channels in Fig \ref{Fig:SLIT RELAXATION PLOT 2}b) occurs since we are plotting attenuation per wavelength (following (\ref{definition: attenuation p wavelength})) and is therefore caused due to the sharp decrease in phase speed observed in this region. Moreover, it is physically useful to also have an idea of this quantity along a fixed distance. In Figure \ref{Fig:per unit metre attenuation} we plot attenuation per unit metre (given by $20 \operatorname{Im}(\Bar{k}) \log_{10}(\me)$) as a function of channel width, and give differences between the viscous/inviscid fluid cases from (\ref{symmetricVE channel :VE-VE-VE}), (\ref{inv-VE 72}) respectively at different values of the Deborah number. In particular, we observe how for narrower slits, the large reduction in phase speed observed above results in shorter wavelengths and consequently higher attenuation per unit metre.

\newpage
\subsection{Limit B): Fluid-loaded plates}\label{Section: Fluid loaded plates}
We now want to explore the reciprocal situation to that just seen above,  namely, Medium 1 in the right of Figure \ref{Fig:Geometries} now corresponds to a solid, whereas Medium 2 is a fluid (water), so that we have fluid-loaded plates. Much like for the half-space, in certain scenarios (hard interfaces) the fluid loading simply acts as a small perturbation to the traditional Lamb DEs for natural modes on a plate in vacuum (i.e.\ stress-free). In these cases the consideration of fluid loading causes the (pure) Lamb mode solutions to become leaky as some of their energy gets shed via radiation of acoustic waves in the fluid, and therefore the resulting modes are commonly addressed as `Leaky--Lamb' modes in the literature, which have been widely studied including the effect of boundary layer losses e.g.\ \cite{zhu1995propagation, nayfeh1997excess, rokhlin1989topology}. 

However, the presence of Material 2 (here a fluid) gives rise to an additional set of solutions of a similar nature to the half-space Scholte--Stoneley addressed in Section \ref{subsubsec: Sch Mode}. Naturally, the difference here is that for thin plates (compared to wavelength) these Sc\ type surface modes become interacting within the plate region and form two coupled interface modes, which are therefore the symmetric/anti-symmetric `coupled plate--Scholte' modes, which we will refer to as `coupled--Scholte' for brevity following \cite{staples2021coupled}. For hard fluid--plate interfaces, the symmetric mode is often ignored since it is non-dispersive and lies on top of the sound line, and therefore the term `quasi--Scholte' mode is often found in the literature to refer only to the anti-symmetric (e.g.\ \cite{cegla2005material}). The situation is nevertheless very different for soft interfaces, and in particular the symmetric coupled Sc can become highly dispersive, deviating from the Sc and fluid velocities, as recently confirmed experimentally in \cite{staples2021coupled}.

Following the same structure that we have employed in the preceding sections, here we want to emphasise the extra physical insights that our framework provides for the behaviour of coupled Sc modes. That is, the effects of the thermo-viscous boundary layers on either side of the plate as well as the TVE of the plate itself, particularly when stress relaxation is considered.

Given the non-dimensionalisation in Section \ref{subsection:non-dimensionalization}, in this limit we let $\bar{c}_\square \equiv \bar{c}_s $ i.e.\ we choose the lossless shear speed of sound in each solid material as appropriate (for steel $\bar{c}_s=3000$ m/s, and for PVC $\bar{c}_s=1100$ m/s) and similarly for the mass density (for steel $\bar{\rho}_1=7871$ kg/m$^3$, and for PVC $\bar{\rho}_1=1360$ kg/m$^3$). With regards to the choice of VE frequency dependence, we will follow a similar structure to what was done in the preceding sections: first we assume the KVM for the solid according to Table \ref{table: VE parameters}, and then we consider the SLSM in  Section \ref{subsection:Plates relaxation}.

\subsubsection{Phase speed and attenuation}
 By the equivalence of the DEs in consideration, in order to find the roots of (\ref{symmetricVE channel :VE-VE-VE}), (\ref{ANTIsymmetricVE channel :VE-VE-VE}) using \textit{fsolve}, we can make use of the same iterative procedure used for the slit, as explained in Section \ref{subsection: slit phase speed / attenuation} so that the initial value in the thick plate limit relies on the roots of (\ref{StoneleyDE}). { Similarly, when thermal effects are considered, for the initial roots of the corresponding DEs (\ref{TVE-TVE-SYM}), (\ref{TVE-TVE-ANTISYM}) we focus on the thick plate limit, given by the TVE Stoneley DE (\ref{stoneley:TVE-TVE}).} As we have been doing thus far, in order to study the influence of the fluid viscosity, it is convenient to consider the inviscid limit such that $\mu_2(\omega) \rightarrow 0$, from which it follows that $k_{\Phi_2}^2 \rightarrow \infty$ and hence $\gamma_{\Phi_2}, B_2 \rightarrow \infty$ whereas $\mathcal{Q}\rightarrow 2  c_{\Phi_1}^2$ so that (\ref{symmetricVE channel :VE-VE-VE}), (\ref{ANTIsymmetricVE channel :VE-VE-VE}) become
\begin{subequations}
\label{Channel :Inv-VE-Inv}
\begin{align} \label{symmetricVE channel :Inv-VE-Inv}
    & \left[\left(2k^2 - k_{\Phi_1}^2 \right)^2 \coth{(\gamma_{\phi_1})} - 4 k^2 \gamma_{\phi_1} \gamma_{\Phi_1} \coth{(\gamma_{\Phi_1})} \right] + \rho_s \frac{\gamma_{\phi_1}  k_{\Phi_1}^4 }{\gamma_{\phi_2} } = 0, \\ \label{ANTIsymmetricVE channel :Inv-VE-Inv}
 & \left[\left(2k^2 - k_{\Phi_1}^2 \right)^2 \tanh{(\gamma_{\phi_1})} - 4 k^2 \gamma_{\phi_1} \gamma_{\Phi_1} \tanh{(\gamma_{\Phi_1})} \right] + \rho_s \frac{\gamma_{\phi_1}  k_{\Phi_1}^4 }{\gamma_{\phi_2} } = 0.
\end{align}
\end{subequations}
for symmetric and anti-symmetric modes respectively, which we will both be considering in this section. From (\ref{Channel :Inv-VE-Inv}), we observe explicitly that the terms in square brackets correspond to the classical Lamb DEs for modes on a free plate (in the absence of any loading $\rho_s=0$), where the VE effects are captured through the frequency varying material parameters appearing in the wavenumbers. The role of the density ratio $\rho_s$ is analogous to (\ref{ScholteDE}) for the half-space. For $\rho_s \ll 1$ we can see why (\ref{Channel :Inv-VE-Inv}) can be treated as a small perturbation to the stress-free Lamb DEs \cite{jia1997modal}, nevertheless as the fluid density approaches that of the solid $\rho_s \rightarrow 1$, the similarities between the spectra can disappear completely \cite{rokhlin1989topology}.

\begin{figure}
\centering
\includegraphics[scale=0.38]{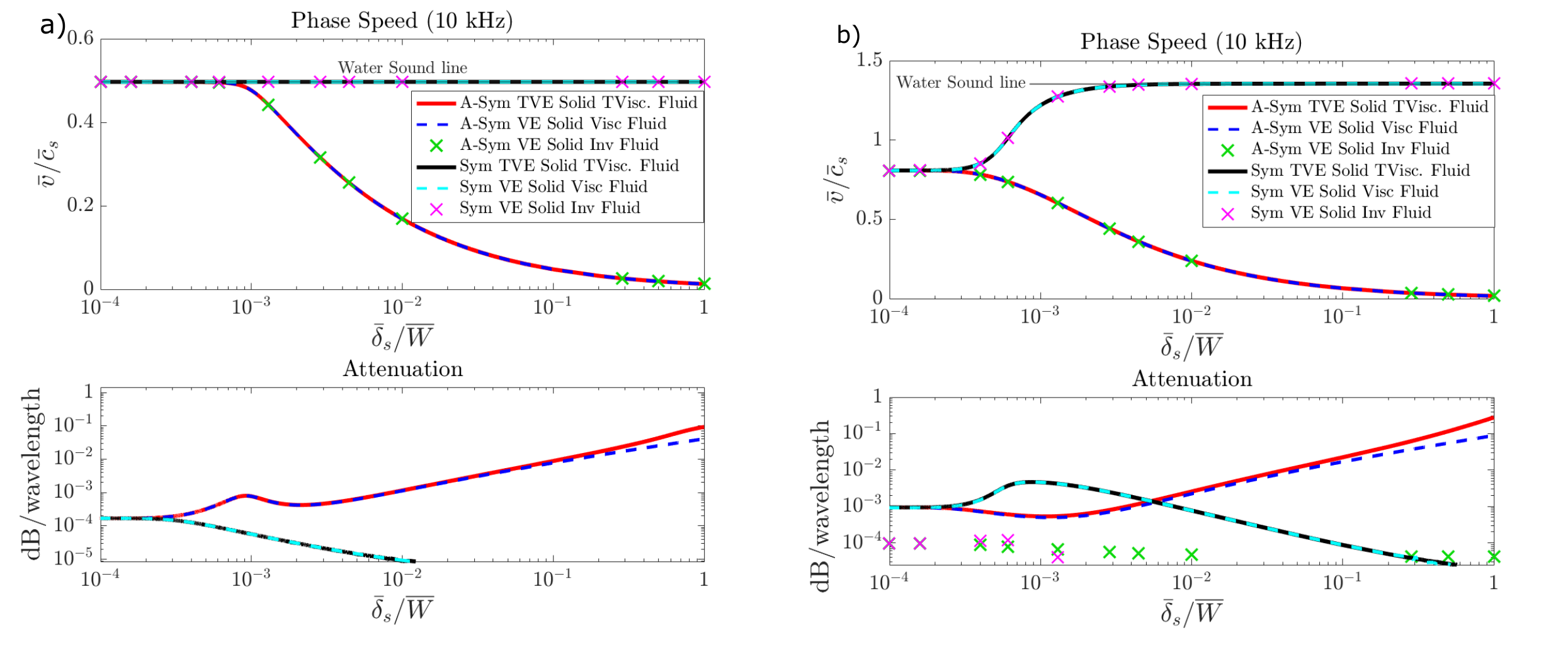}
\caption{Relative phase speed/attenuation at $10$ kHz for coupled Sc modes in water-loaded solid dissipative plates of decreasing width for: a) Steel, and b) PVC, whose material properties are in Tables \ref{table: VE parameters}, \ref{table: Th parameters}. The initial values (thick plate limit) are dictated by the Stoneley DE (\ref{StoneleyDE}) and therefore the initial phase speeds are equivalent to those in Figure \ref{Fig:Phase/Atten Fluid filled channel} (noting the change in non-dimensionalisation).}
\label{Fig:Phase/Atten Coupled Scholte mode}
\end{figure}

In Figure \ref{Fig:Phase/Atten Coupled Scholte mode} we give the relative phase speed/attenuation as a function of the plate thickness in terms of $\bar{\delta}_s/\bar{W}$ at $10$ kHz for: (a) water--steel, and (b) water--PVC. In both cases we observe how, in the thin plate (low frequency) limit the anti-symmetric mode tends to zero, whereas the symmetric mode tends to the fluid's sound speed. Conversely in the thick plate (high frequency) limit, the two curves converge to the Stoneley--Scholte value. These observations are in agreement with the analysis in \cite{osborne1945transmission} and the dispersion curves in \cite{staples2021coupled}. In terms of attenuation, we first notice the significantly smaller magnitude compared to those in the slit e.g.\ Figure \ref{Fig:Phase/Atten Fluid filled channel}, which is expected, since here the boundary layers are located on either side exterior to the plate, and are therefore never interacting (which is the regime with major viscous losses as observed in the preceding slit section). Furthermore, it is also apparent from the low attenuation values that the TVE effects of the plates, with the values used, have small impact. Given this, we nevertheless observe how the viscous boundary layer increases the attenuation of the anti-symmetric mode as the plate becomes thinner in both cases, {with a small albeit notable distinction between the TVE and VE solutions as we approach $\bar{W} = \bar{\delta}_s$ in both solids Figs \ref{Fig:Phase/Atten Coupled Scholte mode}a),b)}. The general increase in attenuation does not occur for the symmetric mode but instead, we observe a local maximum for water--PVC Figure \ref{Fig:Phase/Atten Coupled Scholte mode}b) around $\bar{\delta}_s/\bar{W} \approx 10^{-3}$ which occurs near (but not exactly) the inflection point observed in the mode's phase speed, but is not observed for steel since the symmetric mode's speed remains constant Figure \ref{Fig:Phase/Atten Coupled Scholte mode}a).

\begin{figure}
\centering
\includegraphics[scale=0.38]{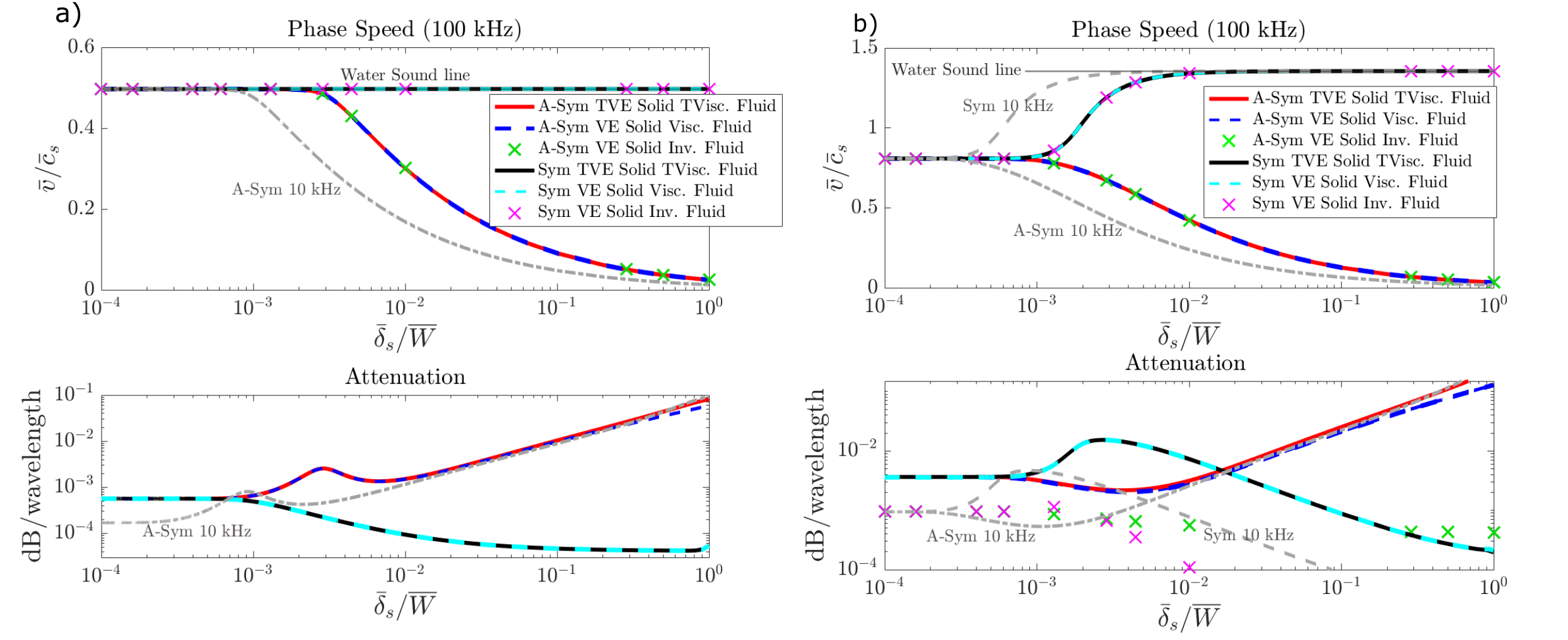}
\caption{Relative phase speed/attenuation at $100$ kHz for coupled Sc modes in water-loaded dissipative solid plates of decreasing width for: a) Steel, and b) PVC, whose material properties are in Tables \ref{table: VE parameters}, \ref{table: Th parameters}. The curves in grey correspond to the TVE Sym/A-Sym Coupled Sc mode for viscous water at $10$ kHz, represented as black and red curves in Figure \ref{Fig:Phase/Atten Coupled Scholte mode}. Except for the phase speed of the Sym mode in a), notable dispersive effects in phase speed and attenuation are observable in both media.}
\label{Fig:Phase/Atten Coupled Scholte mode Hi Freq}
\end{figure}

In order to further analyse the dispersion of these modes, in Figure \ref{Fig:Phase/Atten Coupled Scholte mode Hi Freq} we give an equivalent plot but increase frequency $\omega$ by an order of magnitude (to $100$ kHz) and superimpose (in grey) the TVE results from Figure \ref{Fig:Phase/Atten Coupled Scholte mode} for ease of comparisons. We observe how this results in the separation of the phase speeds (between the symmetric and anti-symmetric modes) at thinner plate thicknesses and correspondingly so does the maximum attenuation of the viscous symmetric mode for PVC in Fig \ref{Fig:Phase/Atten Coupled Scholte mode Hi Freq}b) which also increases in magnitude as to be expected. In general, we observe that the inviscid fluid solutions obtained from (\ref{Channel :Inv-VE-Inv}) have an excellent agreement in phase speed with the full viscous solutions, but naturally cannot predict the attenuation due to the boundary layers.

\subsubsection{Displacement fields}
\begin{figure}
\centering
   \includegraphics[trim={4cm 0 2cm 0},scale=0.43]{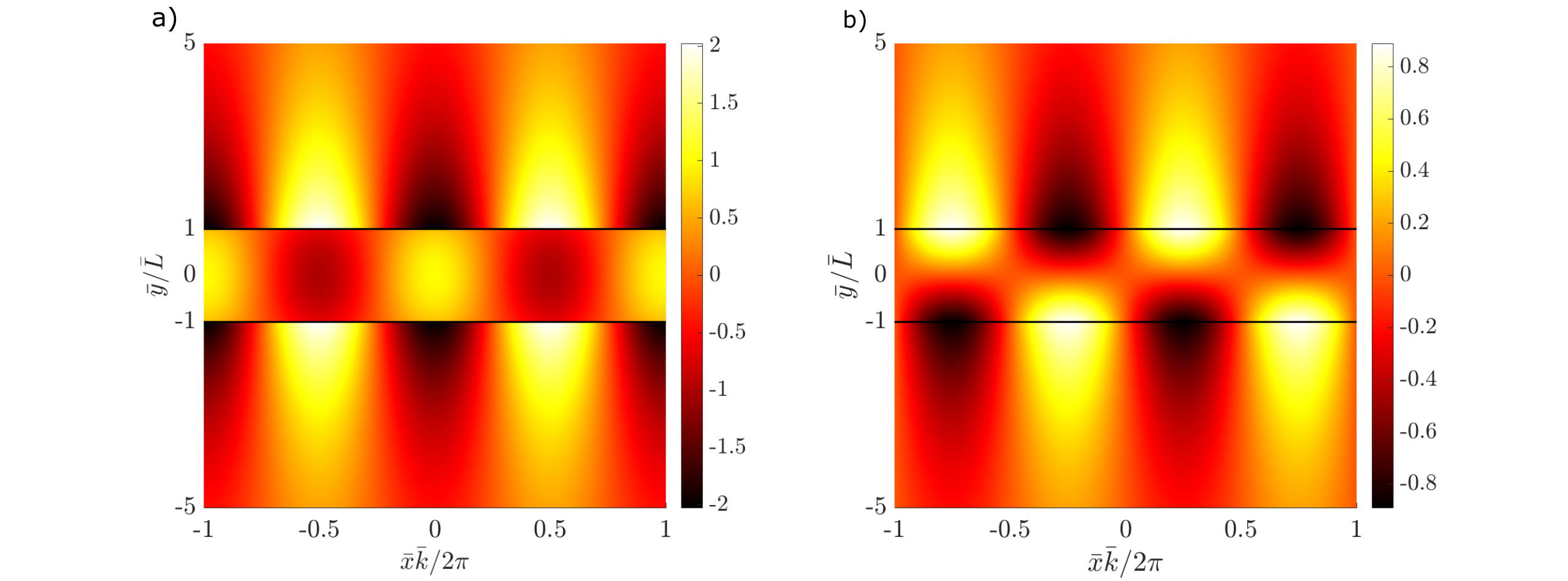}
\caption{Real part of the displacement fields for the symmetric coupled Sc mode propagating in a PVC plate immersed in water at $10$ kHz, for a plate thickness of $\bar{\delta}_s/\bar{W}=10^{-3}$ with values from Table \ref{table: VE parameters} and DE (\ref{symmetricVE channel :VE-VE-VE}). a) represents the horizontal particle displacement $\operatorname{Re}\{u_x(x,y)\}$, and b) the vertical displacement $\operatorname{Re}\{u_y(x,y)\}$. The plots have been normalized such that  $u_x(y=x=0)=1$, and different lengthscales are used for the $x$ and $y$ directions, with the black lines representing the plate boundaries. The difference in the values of the colorbar show that the predominant motion is parallel to the interface.}
\label{Fig:Displacements PVC PLate SYM}
\end{figure}

\begin{figure}
\centering
   \includegraphics[trim={4cm 0 2cm 0},scale=0.43]{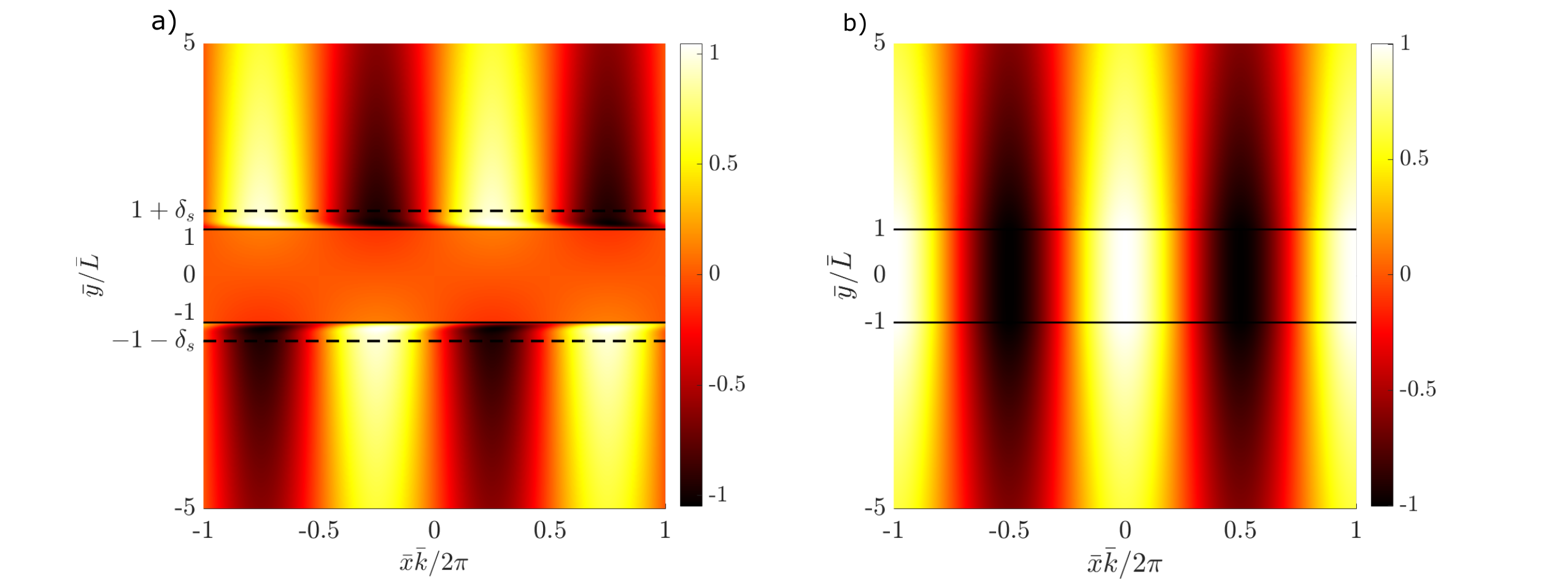}
\caption{Real part of the displacement fields for the anti-symmetric coupled Sc mode propagating in a PVC plate immersed in water at $10$ kHz, for a plate thickness of $\bar{\delta}_s/\bar{W}=0.2$ with values from Table \ref{table: VE parameters} and DE (\ref{ANTIsymmetricVE channel :VE-VE-VE}). a) represents the horizontal particle displacement $\operatorname{Re}\{u_x(x,y)\}$, and b) the vertical displacement $\operatorname{Re}\{u_y(x,y)\}$. The plots have been normalized such that  $u_y(y=x=0)=1$, and different lengthscales are used for the $x$ and $y$ directions with the black lines representing the plate boundaries.}
\label{Fig:Displacements PVC PLate ANTI-SYM}
\end{figure}

The energy density associated with the coupled Sc mode for hard/soft plates  of various thicknesses was analysed respectively in \cite{cegla2005material, staples2021coupled}, so here we will focus on the mode's displacement fields. In Figure \ref{Fig:Displacements PVC PLate SYM} we show heatmaps of (the real part of) the horizontal/vertical particle displacements fields for the symmetric coupled Sc mode in a PVC plate (Table \ref{table: VE parameters}) loaded with water at $10$ kHz for a plate thickness of $\bar{\delta}_s/\bar{W}=10^{-3}$ (see Figure \ref{Fig:Phase/Atten Coupled Scholte mode}b) for reference). From the colorbars we note that the predominant motion is actually parallel to the interface (Figure \ref{Fig:Displacements PVC PLate SYM}a)) with significant motion coupled in both the plate and near the interface in the fluid. The anti-symmetric mode is given in Figure \ref{Fig:Displacements PVC PLate ANTI-SYM}, noting that in this case the magnitude of the motion is much more distributed in both directions, with the predominant motion in the plate being perpendicular to the interface which is expected due to the `bending' nature of the anti-symmetric mode. Note that, for this plot we have used a much thinner plate of $\bar{\delta}_s/\bar{W}=0.2$ in order to further illustrate the viscous boundary layers (Fig \ref{Fig:Displacements PVC PLate ANTI-SYM}a).

\begin{figure}
\centering
\includegraphics[scale=0.4]{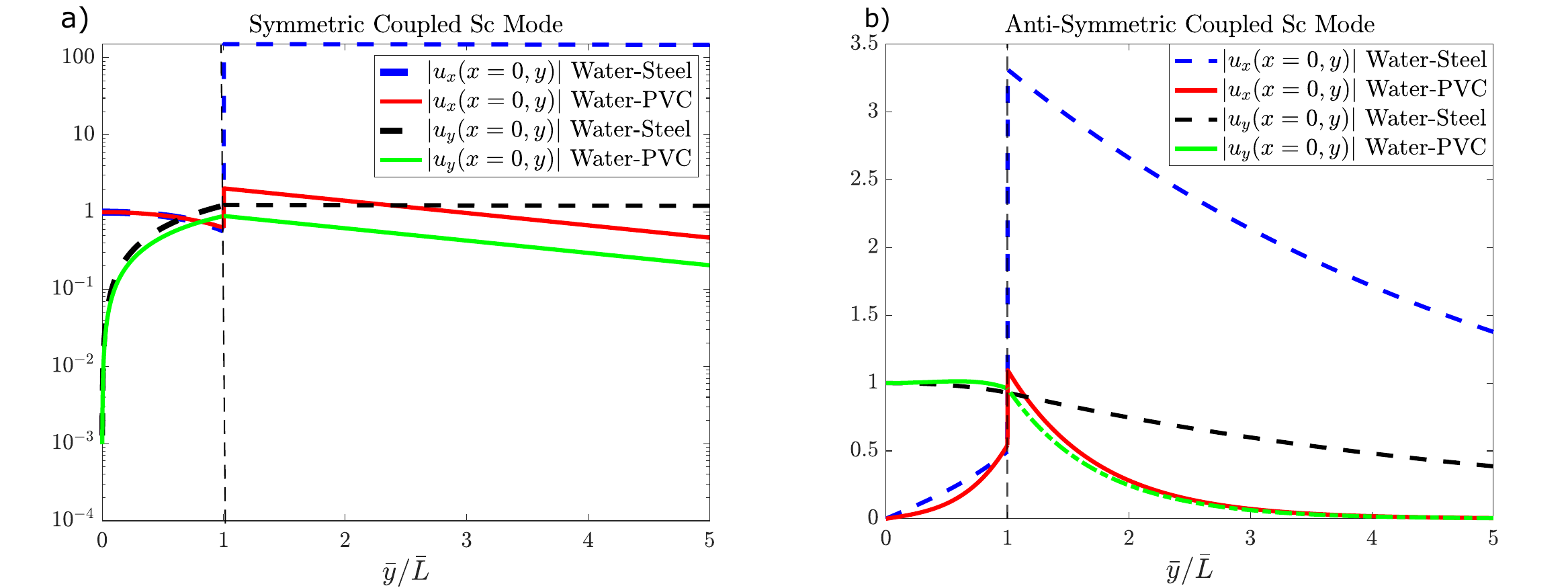}
\caption{Comparisons between the magnitude of the components of the displacement fields of steel/PVC plates loaded with water at at $10$ kHz for the sym. coupled Sc mode (a), and the anti-sym. coupled Sc mode (b) normalized in each case s.t. the displacement at the center of the plate is $1$. The plate thickness is $\bar{\delta}_s/\bar{W}=10^{-3}$ with values from Table \ref{table: VE parameters}. The apparent discontinuities for $u_x$ at the plate interface $y=1$ are simply due to the thinness of the boundary layer region with the large $y$ scale employed.}
\label{Fig:Plate Displacement comparisons}
\end{figure}

In Figure \ref{Fig:Plate Displacement comparisons} we give direct comparisons of (the absolute value of) the particle displacements as seen in Figs \ref{Fig:Displacements PVC PLate SYM}, \ref{Fig:Displacements PVC PLate ANTI-SYM} for $y \geq 0$ evaluated at $x=0$, as well as the equivalent results for a steel water-loaded plate (all curves are solutions to the VE DEs (\ref{symmetricVE channel :VE-VE-VE}), (\ref{ANTIsymmetricVE channel :VE-VE-VE})). From Fig \ref{Fig:Plate Displacement comparisons}a) we first observe how indeed for the symmetric mode the parallel particle displacement $u_x$ is dominant over the perpendicular $u_y$. For PVC we observe a decay within the fluid region, as opposed to water--steel, where the majority of the motion (hence energy) lies in the fluid region, with no decay observed. This situation resembles the half-space observations e.g.\ Figure \ref{Fig:Stoneley/Scholte}d), where we essentially have compressional waves at grazing incidence in the fluid region travelling at the sound speed (Fig \ref{Fig:Phase/Atten Coupled Scholte mode}a)), with little coupling in the plate region. For the anti-symmetric mode (Fig \ref{Fig:Plate Displacement comparisons}b) in water--PVC, we note the more rapid decay of the motion within the fluid, whereas for water--steel we again observe a significantly larger magnitude of the horizontal particle displacement in the fluid region, although here the decay is noticed, in agreement with \cite{cegla2005material}. Note that the apparent discontinuities for $u_x$ at the plate interface $\bar{y}=\bar{L}$ are simply due to the extremely thin boundary layer regions (as in Fig \ref{Fig:Displacements fluid filled channels}a)) with the chosen parameters ($\bar{\delta}_s/\bar{W}=10^{-3}$ at $10$ kHz) which make them not visible at the provided $y$ scale.


\subsubsection{Influence of stress relaxation}\label{subsection:Plates relaxation}
Finally, we want to pay attention to the effect of stress relaxation on the soft plate and observe the corresponding effects on the phase speed/attenuation of the coupled Scholte mode. As we have done in Sections \ref{subsection: HS stress relaxation}, \ref{subsection: Slit stress relaxation} we proceed by letting the Young's modulus of the plate $E_1(\omega)$ follow the SLSM whilst its Poisson's ratio $\nu_1$ remains constant, as in (\ref{Young's Mod SLSM}) (with subscripts in the moduli interchanged from `2' to `1') noting that $t_r=\bar{c}_s \bar{t}_r/\Bar{L}$. In the illustrations below we also let $\bar{E}_\infty=4.4524$ GPa so that the rubbery phase limit of the PVC material corresponds to the (fixed) value from Table \ref{table: VE parameters} (as we did for the slits in Section \ref{subsection: Slit stress relaxation}) {and ignore thermal effects.} 

\begin{figure}
\centering
\includegraphics[scale=0.4]{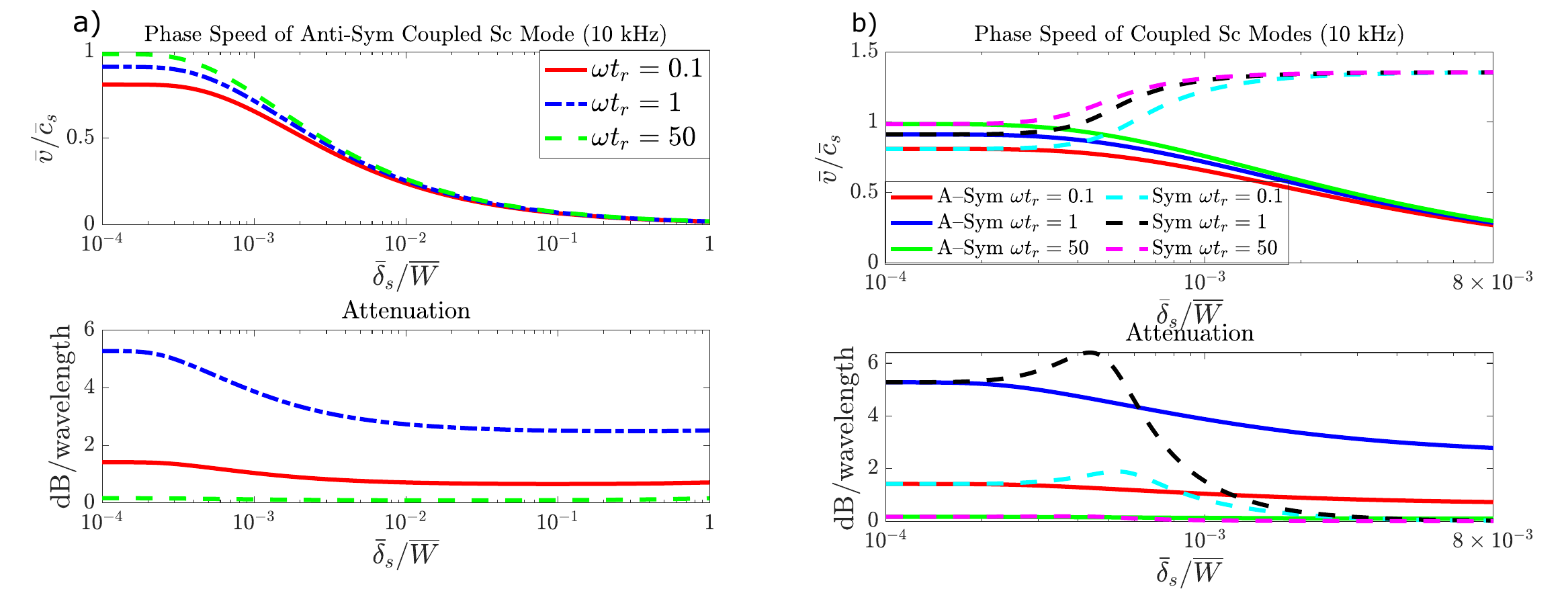}
\caption{Relative phase speed/attenuation at 10 kHz for coupled Sc modes in water-loaded VE PVC plates of decreasing width according to the SLSM with $\bar{E}_0/\bar{E}_\infty=1.58$, whilst keeping the Deborah number fixed (and in all cases $\Bar{c}_s=1100$ m/s). The fluid is assumed to be viscous so that all the curves correspond to roots of (\ref{symmetricVE channel :VE-VE-VE}) and (\ref{ANTIsymmetricVE channel :VE-VE-VE}). In a) we only show the anti-symmetric mode for a very wide range of plate widths, whereas in b) we also include the symmetric mode for plate thicknesses in the region of interest. In particular, we note the enhancement of the global attenuation maxima for the symmetric mode which subsequently tends to zero as the mode becomes non-dispersive.}
\label{Fig:Plate Relaxation}
\end{figure}

In Figure \ref{Fig:Plate Relaxation} we give the phase speed/attenuation of the symmetric and anti-symmetric coupled Sc mode as a function of plate thickness (as in Fig \ref{Fig:Phase/Atten Coupled Scholte mode}) with fixed Deborah numbers $\omega t_r=0.1, 1, 50$ so that we cover the rubbery and glassy phases, as well as glass transition. Furthermore, in this figure we have chosen a moderate value of the ratio $\bar{E}_0/\bar{E}_\infty=1.58$, see Figure \ref{Fig:Phase/Atten Relaxation HS}b). For the anti-symmetric mode (Figure \ref{Fig:Plate Relaxation}a)) we observe how the initial phase speed values dictated by the Stoneley DE (\ref{StoneleyDE}) become higher when the material is in the glassy phase, which then tend to zero in the thin plate limit following the observations in Figure \ref{Fig:Phase/Atten Coupled Scholte mode}. In terms of attenuation, we first observe the remarkably higher values for $\omega t_r=0.1, 1$ (at all plate widths) when compared to the KVM results from Figure \ref{Fig:Phase/Atten Coupled Scholte mode}. For the anti-symmetric mode (Figure \ref{Fig:Plate Relaxation}a)) we observe that the attenuation (per wavelength) approaches non-zero values (except for $\omega t_r \gg 1$) in the thin plate limit. With regards to the symmetric mode (Figure \ref{Fig:Plate Relaxation}b)) we first  note that the initial behaviour in both phase speed and attenuation is identical to that of the anti-symmetric mode, as expected following our discussion above.
As the plate width decreases, the mode first enters the dispersive region and continues to asymptote towards the fluid's phase speed $\bar{v}/\bar{c}_s=1.35$ following the observations from Figure \ref{Fig:Phase/Atten Coupled Scholte mode}. When it comes to the attenuation, the global maximum discussed above become significantly enhanced in the dispersive region (particularly for $\omega t_r \approx 1$), after which it monotonically decreases towards zero as the root approaches the fluid's (constant) bulk mode. 

It is worth stressing that this global maximum for the symmetric coupled Sc only arises when the mode's phase speed is dispersive which (for a fixed fluid) requires soft media, so that $\rho_s \approx 1$. For this reason we did not observe it for water-loaded steel plates in Figure \ref{Fig:Phase/Atten Coupled Scholte mode}a), {noting that is the main focus of existing literature \cite{staples2021coupled}}. Some further tests (not shown) confirmed that the inclusion of fluid viscosity has little influence on the presence of this maximum, so that solutions to (\ref{Channel :Inv-VE-Inv}a) (corresponding to $\eta_{\mu_2}=0$) accurately capture this global maximum (as expected given the little magnitude of boundary layer attenuation in Figure \ref{Fig:Phase/Atten Coupled Scholte mode}). {This can also be noticed by the VE Solid--Inviscid fluid results (pink crosses) in the bottom of Fig \ref{Fig:Phase/Atten Coupled Scholte mode Hi Freq}b), albeit smaller due to the absence of fluid viscosity, the local maximum can still be appreciated.}

{Perhaps surprisingly, we have not been able to find discussions about this feature in the existing literature.}


\section{Conclusions}\label{section:conclusions}

{This paper has focused on the influence of thermo-viscous damping in some of the principal modes of wave propagation in two physical systems, namely fluid-filled slits embedded in infinitely extending solids, and fluid-loaded plates, where the fluid considered is water.} These two settings are connected by the fact that the respective waves in the short wavelength limit (with respect to channel/plate width) are governed by the same dispersion equation, namely the Scholte-Stoneley DE. The direct correspondence between the governing equations for thermo-visco-acoustic fluids and thermo-visco-elasticity developed in \cite{garcia2022unified} is used in order to obtain generalised dispersion relations for symmetric and anti-symmetric modes that can govern the two specific set-ups of interest when taking the appropriate limits, which is a particularly convenient aspect of this study. 

For fluid-filled channels, the results are presented as an extension of earlier work by some of the present authors \cite{cotterill2018thermo}. We firstly show how the consideration of soft media requires paying more attention to the way the roots are initially found, as a result of the Scholte mode's phase speed reduction (from the fluid sound line) in water-(soft) solid interfaces. We then emphasise how viscoelastic mechanisms in the solid affect the overall attenuation at {different frequency ranges and channel widths}, for both the commonly used Kelvin-Voigt model (KVM) and the standard linear solid model (SLSM), which incorporates stress relaxation. It has been shown how stress relaxation can greatly affect the phase speed and attenuation of the mode, and in particular the ability to damp energy if the mode can be excited around the glass transition of the material which is characterized by a Deborah number of one, i.e.\ $\omega t_r=1$. {As a result of including thermal coupling, we are also able to justify the claim made in \cite{cotterill2018thermo} that for water-filled slits, thermal effects do not add any significant changes in the regions considered.}

For fluid-loaded plates, {the work here builds on the work in \cite{staples2021coupled}}. Utilizing the same root finding technique as for the channels, where the initial value is based on the Scholte-Stoneley DE behaviour, the dispersion of the phase speed of both the symmetric and anti-symmetric modes as a function of plate thickness is analysed, as well the associated displacement fields. We originally observe very little dissipation compared to the slit case as a result of the boundary layers being localized on the exterior sides of the plate and therefore never interacting regardless of the thinness of the plate. Nevertheless, the presence of a global maximum in the attenuation of the symmetric coupled Scholte mode is noticed for the soft solid only, indicating that it is closely linked to the dispersion in phase speed, and therefore not observed for hard water-solid interfaces. The attenuation increases significantly for both modes when stress relaxation is included in the soft plate, and in the thin-plate limit the attenuation of anti-symmetric modes tends to constant non-zero values for each $\omega t_r$, whereas the dissipation of the symmetric mode becomes zero as soon as the phase speed becomes constant. Nevertheless, the global maximum in attenuation observed previously is largely enhanced, particularly around glass transition. {We believe that this global maximum should be observable experimentally following the work in \cite{staples2021coupled} and perhaps can even be interesting to the non-destructive evaluation/testing community for possible applications.}

More generally and perhaps critically, we find that although the widely used KVM (even with varying viscosity coefficient to match experimental results) can accurately predict the loss at particular frequencies as shown in e.g.\ \cite{favretto1997excitation}, with such models the dispersion of the real parts of these elastic moduli (and hence phase speed of the associated modes) cannot be captured. We show this can be crucially important, particularly at frequencies near the glass transition. We further note that the SLSM used here to capture stress relaxation, namely Prony series with a single relaxation time, is the simplest model that captures long time solid-like `glassy' behaviour and generally, the VE behaviour of polymers is much more complex \cite{sharpe2008springer, liao2006estimation}. Furthermore, the relaxation results are based around a PVC sample with $\bar{E}_\infty = 4.4524$ GPa, such that the half-space Scholte mode propagates at a phase speed approximately $60$\% of that of water and is therefore considered soft in this context, especially when compared to the steel sample analysed here, for which the Scholte mode propagates essentially at the speed of sound of water. Nevertheless, the amplitude ratio $\bar{E}_0/\bar{E}_\infty$ is well known to become particularly large for significantly softer media, although in terms of practicality this is offset by the fact that in this regime the Scholte mode is likely to be much more difficult to excite. {These results strongly  motivate the need for further experimental results in the frequencies of interest in order to be able to ascertain under which conditions these mechanisms can be exploited, as well as relevant data for relaxation times and  short/long time moduli.}

\section*{REFERENCES}
\begingroup
\renewcommand{\section}[2]{}

\bibliographystyle{unsrt}

\endgroup

\begin{appendices}

\section{Inclusion of thermal expansion effects}\label{appendix: TVE}
Despite a priori not expecting to add major differences to the physical results obtained with the VE theory presented in the main text since water is the only fluid media used in the calculations, we discuss how to obtain the more general thermo-visco-elastic (TVE) equivalent equations from the governing equations. 
\begin{table}[H]
\centering
\small
\begin{tabular}{ |p{4.5cm}||p{1.85cm}|p{1.1cm}|p{1.3cm}|p{1.3cm}|p{1.15cm}|p{1.15cm}|}
 \hline
 \multicolumn{7}{|c|}{ {\textbf{Thermal Parameters}}} \\
 \hline
 \hspace{1.5cm} Parameter& Unit & Symbol  & Water & Steel & PVC & $M_{\tau=0.9}$ \\
 \hline
 Ratio of specific heats &  -- & $\gamma$  & 1.0095  & 1.0034 & 1.05 & 1.048\\
Specific heat at constant volume &  [J kg$^{-1}$K$^{-1}$] & $c_v$  & 4188  & 498.325 & 1238.09 & 1164.11\\
Ambient temperature &  [K] & $T_0$  & 283.16  & 293 & 300 & 299.3 \\
 Thermal conductivity &  [W m$^{-1}$K$^{-1}$]  & $\mathscr{K}$  & 0.597 & 30 & 2 & 4.8\\
 Coefficient of thermal expansion & [K$^{-1}$] & $\alpha$  & 8.16$\times 10^{-5}$  & 1.7$\times 10^{-5}$ & 3.5$\times 10^{-4}$ &2.2$\times 10^{-4}$ \\
 \hline
\end{tabular}
\caption{Thermal parameter values used for water, steel and PVC employed in the TVE calculations throughout. Recall that $M_{\tau=0.9}$ is a made-up material defined in (\ref{linear transition steel->PVC}).}
\label{table: Th parameters}
\end{table}

\subsection{Governing equations}

Following \cite{garcia2022unified} we have that, when taking into account strain histories but ignoring corresponding thermal histories, a Helmholtz free energy can be derived (from a state of zero strain $\hat{\bm{\varepsilon}}=\mathbf{0}$ and constant temperature ${T}={T}_0$) from which the stress strain relations and entropy follow, namely
\begin{subequations}\label{TVE Non Local Eqns}
\begin{align}
    \hat{\bm{\sigma}} &= \int_{- \infty}^{{t}} 2 \hat{\mu}({t}-{\mathcal{T}})
    \frac{\partial \hat{\bm{e}}(\mathcal{T})}{\partial \mathcal{T}} \diff {\mathcal{T}} + \left(\int_{- \infty}^{{t}}  \hat{K}({t}-{\mathcal{T}}) \tr \left(\frac{\partial \hat{\bm{\varepsilon}}(\mathcal{T})}{\partial \mathcal{T}} \right) \diff {\mathcal{T}} - \alpha {T}_0 K \hat{\theta} \right) \bm{I},\\ \label{entropy} 
    \hat{\mathfrak{s}} &= {\mathfrak{s}}_0 + {c}_v \hat{\theta} +\frac{{\alpha} {K}}{{\rho}_0} \tr (\hat{\bm{\varepsilon}}),
\end{align}
\end{subequations}
where ${T}_0$, $\hat{\theta}=({T}-{T}_0)/{T}_0$ represent the background temperature and non-dimensional temperature, $\alpha$ is the coefficient of thermal expansion, $\hat{\mathfrak{s}}$ denotes the entropy per unit mass and $c_v$ is the specific heat at constant volume (strain) which is related to the specific heat at constant pressure (stress) $c_p$ by $c_p/c_v = \gamma$ where $\gamma$ denotes the ratio of specific heats. Naturally, we must still satisfy the momentum equations (\ref{eqn: of motion time domain}) together with the energy balance equation, which is given by
\begin{equation}
    \label{eqn:IsotropicEnergyChristensen}
    \mathscr{K} \Delta \hat{\theta} - {\rho}_0 {c}_v \frac{\partial \hat{\theta}}{\partial t}  = \alpha K \tr \left(\frac{\partial \hat{\bm{\varepsilon}}(\mathcal{T})}{\partial \mathcal{T}} \right),
\end{equation}
where $\mathscr{K}$ is the thermal conductivity (the units of these newly introduced thermal quantities are summarized in Table \ref{table: Th parameters}). Assuming now that the fields are time-harmonic, of the form
\begin{equation}\label{eqn:DevStressTimeHarmonic}
    \{\hat{\mathbf{u}},\hat{\theta},\hat{\bm{\sigma}},\hat{\bm{\varepsilon}},\hat{\bm{e}} \}({\mathbf{x},t)}= \operatorname{Re}{\{ \{ \mathbf{u},\theta,\bm{\sigma},\bm{\varepsilon},\bm{e} \}({\mathbf{x})}\me^{-\mi {\omega} {t}}\}},
\end{equation}
we show in detail in \cite{garcia2022unified} that introducing the wave potentials in the Helmholtz form
\begin{align}\label{eqn:diplacementDecompos}
    {{\mathbf{u}}} &= {\mathbf{\nabla}} \left({\varphi} + \vartheta \right) + {\mathbf{\nabla}} \times  {\mathbf{\Phi}}, & {\nabla} \cdot {\mathbf{\Phi}} &= 0,
\end{align}
the governing equations (\ref{eqn: of motion time domain}), (\ref{eqn:IsotropicEnergyChristensen}) reduce to solving
\begin{align}\label{TVE potential equations}
 \Delta {\vartheta} +{k}_\vartheta^2 {\vartheta}=0, \quad
    \Delta {\varphi}+{k}_\varphi^2 {\varphi}=0, \quad  \Delta {\mathbf{\Phi}} + {k}_\Phi^2 {\mathbf{\Phi}} =\mathbf{0},
\end{align}
where the thermo-compressional wavenumbers can be written as
\begin{equation}
    {k}_\vartheta^2 = {a} + {b}, \quad {k}_\varphi^2 = {a} - {b}, \qquad {a}=\frac{1}{2}\left({k}_\theta^2+{k}_\phi^2-L_{\theta}{L}_{\phi} \right), \quad \text{and} \quad {b} =\sqrt{{a}^2-{k}_\phi^2{k}_\theta^2},
\end{equation}
and
\begin{equation}
       {k}_{\theta}^2=\mi {c}_v\frac{{\rho}_0 {\omega}}{{\mathscr{K}}}, \quad {k}_{\phi}^2=\frac{{\rho}_0 {\omega}^2}{{\hat{K}(\omega)}+\frac{4}{3}{\hat{\mu}(\omega)}}, \quad {k}_{\Phi}^2=\frac{{\rho}_0 {\omega}^2}{{\hat{\mu}(\omega)}}, \quad {L}_\phi=\frac{\mi {\alpha} {K} {\omega}}{{\mathscr{K}}},  \quad L_\theta=-\frac{{\alpha}  {T}_0 {K}}{{\hat{\lambda}(\omega)}+2{\hat{\mu}(\omega)}}.
\end{equation}
Furthermore, the corresponding temperature difference field is given by 
$\theta = \mathscr{T}_{\varphi} \varphi + \mathscr{T}_{\vartheta} \vartheta$, where $\mathscr{T}_{\varphi} = (k_\varphi^2 - k_\phi^2)/L_\theta$ and similarly $\mathscr{T}_{\vartheta} = (k_\vartheta^2 - k_\phi^2)/L_\theta$.

\subsection{The TVE Stoneley-Scholte Dispersion Equation}\label{appendix:section: TVE HS}
Given the generalised equations in the TVE context presented above (\ref{eqn:diplacementDecompos}), (\ref{TVE potential equations}) the potentials in (\ref{Potentials:Stoneley}) are now written as 

\begin{subequations}\label{Appendix:Potentials:Stoneley}
\begin{align} \label{Appendix:Stoneley Compressional}
\bar{\varphi}_1 &= \bar{P}_1 \me^{ - \bar{\gamma}_{\varphi_1} \bar{y} + \mi \bar{k} \bar{x}},  
& \bar{\varphi}_2 &= \bar{P}_2\me^{  \bar{\gamma}_{\varphi_2} \bar{y} + \mi \bar{k} \bar{x}},\\  \label{Appendix:Stoneley Therm}
\bar{\vartheta}_1 &= \bar{\mathcal{T}}_1 \me^{ - \bar{\gamma}_{\vartheta_1} \bar{y} + \mi \bar{k} \bar{x}},  
& \bar{\vartheta}_2 &= \bar{\mathcal{T}}_2\me^{  \bar{\gamma}_{\vartheta_2} \bar{y} + \mi \bar{k} \bar{x}},\\  \label{Appendix:Stoneley Shear}
\bar{\Phi}_1 &= \bar{S}_1 \me^{ - \bar{\gamma}_{\Phi_1} \bar{y} + \mi \bar{k} \bar{x}},  &
\bar{\Phi}_2 &= \bar{S}_2 \me^{ \bar{\gamma}_{\Phi_2} \bar{y} + \mi \bar{k} \bar{x}}, 
\end{align}
\end{subequations}
where $\bar{\gamma}_{\Phi_1}$, $\bar{\gamma}_{\Phi_2}$ remain identical to those in the isothermal case (\ref{defn: sqrt functions HS}) whereas
\begin{equation}\label{Appendix:defn: sqrt functions HS}
    \bar{\gamma}_{\varphi_1} = (\bar{k}^2 - \bar{k}_{\varphi_1}^2)^{1/2}, \quad \bar{\gamma}_{\varphi_2} = (\bar{k}^2 - \bar{k}_{\varphi_2}^2)^{1/2}, \quad \bar{\gamma}_{\vartheta_1} = (\bar{k}^2 - \bar{k}_{\vartheta_1}^2)^{1/2}, \quad \bar{\gamma}_{\vartheta_2} = (\bar{k}^2 - \bar{k}_{\vartheta_2}^2)^{1/2},
\end{equation}
 for some complex amplitudes $\bar{P}_1,\bar{P}_2,\bar{\mathcal{T}}_1,\bar{\mathcal{T}}_2,\bar{S}_1,\bar{S}_2$. As discussed above, we must ensure that the choice of the various branch cuts of the square root functions in (\ref{Appendix:defn: sqrt functions HS}) is consistent with causality. At the interface between the two media, the solutions must again satisfy continuity of traction and displacement, and in addition, continuity of temperature and temperature flux BCs, that is on $\bar{y}=0$,
\begin{equation}\label{TVE 2 HS BCs 1}
    \bar{\sigma}_{yy}^1 = \bar{\sigma}_{yy}^2, \;\;  
    \bar{\sigma}_{xy}^1 = \bar{\sigma}_{xy}^2, \;\;
    \bar{\mathbf{u}}_1 = \bar{\mathbf{u}}_2,  \;\; {\theta}_1 = {\theta}_2, \;\;
    \bar{\mathscr{K}}_1 \bar{\nabla} {\theta}_1 \cdot \mathbf{e_y} = \bar{\mathscr{K}}_2 \bar{\nabla} {\theta}_2 \cdot \mathbf{e_y}.
\end{equation}
 Substitution of (\ref{Appendix:Potentials:Stoneley}) into (\ref{TVE 2 HS BCs 1}) then yields a homogenous linear system, whose non-trivial solutions are given by
 \begin{equation}\label{stoneley:TVE-TVE}
      \det \bar{\bm{A}} = 0,
 \end{equation}
 where
 \begin{align*} \label{TVE HS Matrix}
    \bar{\bm{A}} = & \left(
\begin{array}{ccc}
 -2 \mi {\bar{\gamma}_{\varphi_1}} \bar{k} {\bar{\mu}_1} & -2 \mi {\bar{\gamma}_{\vartheta_1}} \bar{k} {\bar{\mu}_1} &
   {\bar{\mu}_1} \left(2 \bar{k}^2-{\bar{k}^2_{\Phi_1}}\right)  \\
 2 \bar{k}^2 {\bar{\mu}_1}-{\bar{\alpha}_1 \bar{T}_1} {\bar{K}_1} {\bar{\mathscr{T}}_{\varphi_1}}-{\bar{k}^2_{\varphi _1}}
   \bar{\Gamma}_1 & 2 \bar{k}^2 {\bar{\mu}_1}-{\bar{\alpha}_1 \bar{T}_1} {\bar{K}_1}
   {\bar{\mathscr{T}}_{\vartheta_1}}-{\bar{k}^2_{\vartheta_1}} \bar{\Gamma}_1 & 2 \mi
   {\bar{\gamma}_{\Phi_1}} \bar{k} {\bar{\mu}_1} \\
 \mi \bar{k} & \mi \bar{k} & -{\bar{\gamma}_{\Phi_1}} \\
 -{\bar{\gamma}_{\varphi_1}} & -{\bar{\gamma}_{\vartheta_1}} & -\mi \bar{k} \\
 {\bar{\mathscr{T}}_{\varphi_1}} & {\bar{\mathscr{T}}_{\vartheta_1}} & 0
   \\
 {\bar{\gamma}_{\varphi_1}} {\bar{\mathscr{K}}_1} {\bar{\mathscr{T}}_{\varphi_1}} & {\bar{\gamma}_{\vartheta_1}}
   {\bar{\mathscr{K}}_1} {\bar{\mathscr{T}}_{\vartheta_1}} & 0 \\
\end{array} \numberthis
\right.\\ & \qquad\qquad \left.
\begin{array}{ccc}
 -2 \mi {\bar{\gamma}_{ \varphi_2}} \bar{k} {\bar{\mu}_2} & -2 \mi {\bar{\gamma}_{\vartheta_2}} \bar{k} {\bar{\mu}_2} & -{\bar{\mu}_2} \left(2
   \bar{k}^2 - {\bar{k}^2_{\Phi_2}}\right) \\
   -2 \bar{k}^2 {\bar{\mu}_2}+{\bar{\alpha}_2 \bar{T}_2} {\bar{K}_2}
   {\bar{\mathscr{T}}_{\varphi_2}}+{\bar{k}^2_{\varphi_2}} \bar{\Gamma}_2 & -2 \bar{k}^2 {\bar{\mu}_2}+{\bar{\alpha}_2 \bar{T}_2} {\bar{K}_2} {\bar{\mathscr{T}}_{\vartheta_2}}+{\bar{k}^2_{\vartheta_2}} \bar{\Gamma}_2 & 2 \mi {\bar{\gamma}_{\Phi_2}} \bar{k} {\bar{\mu}_2} \\
   -\mi \bar{k} & -\mi \bar{k} & -{\bar{\gamma}_{\Phi_2}} \\
    -{\bar{\gamma}_{\varphi_2}} &
   -{\bar{\gamma}_{\vartheta_2}} & \mi \bar{k} \\
    -{\bar{\mathscr{T}}_{\varphi_2}} & -{\bar{\mathscr{T}}_{\vartheta_2}} & 0 \\  {\bar{\gamma}_{\varphi_2}} {\bar{\mathscr{K}}_2}
   {\bar{\mathscr{T}}_{\varphi_2}} & {\bar{\gamma}_{\vartheta_2}} {\bar{\mathscr{K}}_2} {\bar{\mathscr{T}}_{\vartheta_2}} & 0 \\
\end{array} \right), 
 \end{align*}
 where $\bar{\Gamma}_i =\bar{\Gamma}_i(\omega) = \hat{\bar{K}}_i(\bar{\omega}) + 4\hat{\bar{\mu}}_i(\bar{\omega})/3$ for $i=1,2$. In the limit $\bar{\alpha}_1,\bar{\alpha}_2 \rightarrow 0$ the thermal and mechanical effects fully decouple and equation (\ref{stoneley:TVE-TVE}) reduces to (\ref{StoneleyDE}) after relabeling.

\subsection{The TVE Slit-Plate Dispersion Equations}\label{appendix:section: TVE Slit/Plate}
For symmetric modes, the generalisation to (\ref{Symm. Waveguides}) in this setting is given by


\begin{subequations} \label{TVE Symm. Waveguides}
\begin{align} \label{TVE Symm. Waveguide Compressional}
\bar{\varphi}_1 &= \bar{\text{P}1}_\text{S} \cosh{(\bar{\gamma}_{\varphi_1} \bar{y})} \me^{\mi \bar{k} \bar{x}},  &
 \bar{\varphi}_2(\bar{x},\bar{y}) = 
\begin{cases}
		  \bar{\text{P}2}_\text{S}\me^{  \bar{\gamma}_{\varphi_2} (\bar{y}+\bar{L}) + \mi \bar{k} \bar{x}}, \qquad & \bar{y} \leq -\bar{L}\\
          \bar{\text{P}2}_\text{S}\me^{  -\bar{\gamma}_{\varphi_2} (\bar{y}-\bar{L}) + \mi \bar{k} \bar{x}}, \qquad & \bar{y} \geq  \bar{L}
\end{cases},\\ \label{TVE Symm. Waveguide THERMAL}
\bar{\vartheta}_1 &= {\text{T}1}_\text{S} \cosh{(\bar{\gamma}_{\vartheta_1} \bar{y})} \me^{\mi \bar{k} \bar{x}},  &
 \bar{\vartheta}_2(\bar{x},\bar{y}) = 
\begin{cases}
		 \bar{\text{T}2}_\text{S}\me^{  \bar{\gamma}_{\vartheta_2} (\bar{y}+\bar{L}) + \mi \bar{k} \bar{x}}, \qquad & \bar{y} \leq -\bar{L}\\
          \bar{\text{T}2}_\text{S}\me^{  -\bar{\gamma}_{\vartheta_2} (\bar{y}-\bar{L}) + \mi \bar{k} \bar{x}}, \qquad & \bar{y} \geq  \bar{L}
\end{cases},\\ \label{TVE Symm. Waveguide Shear}
\bar{\Phi}_1 &= \bar{\text{S}1}_\text{S} \sinh{(\bar{\gamma}_{\Phi_1} \bar{y})} \me^{\mi \bar{k} \bar{x}},  &
\bar{\Phi}_2(\bar{x},\bar{y}) = 
\begin{cases}
		  \bar{\text{S}2}_\text{S}\me^{  \bar{\gamma}_{\Phi_2} (\bar{y}+\bar{L}) + \mi \bar{k} \bar{x}}, \qquad & \bar{y} \leq -\bar{L}\\
          \bar{\text{S}2}_\text{S}\me^{  -\bar{\gamma}_{\Phi_2} (\bar{y}-\bar{L}) + \mi \bar{k} \bar{x}}, \qquad & \bar{y} \geq  \bar{L}
\end{cases},
\end{align}
\end{subequations}
for some complex valued amplitudes $\bar{\text{P}1}_\text{S}, \bar{\text{P}2}_\text{S},\bar{\text{T}1}_\text{S}, \bar{\text{T}2}_\text{S}, \bar{\text{S}1}_\text{S},  \bar{\text{S}2}_\text{S}$. In this case, the continuity conditions (\ref{TVE 2 HS BCs 1}) must be satisfied at the two interfaces between the two media $\bar{y}=\pm \bar{L}$. Substitution of (\ref{TVE Symm. Waveguides}) into (\ref{TVE 2 HS BCs 1}) then yields an equation of the form $\det \bar{\bm{A}}_{Sym} = 0$, where $\bar{\bm{A}}_{Sym}$ is given by

\begin{align*}
     & \left(
\begin{array}{ccc}
 2 \mi {\bar{\gamma}_{\varphi_1}} \bar{k} {\bar{\mu}_1} \sinh{(\bar{\gamma}_{\varphi_1} \bar{L})} & 2 \mi {\bar{\gamma}_{\vartheta_1}} \bar{k} {\bar{\mu}_1} \sinh{(\bar{\gamma}_{\vartheta_1} \bar{L})}&
   {\bar{\mu}_1} \left(2 \bar{k}^2-{\bar{k}^2_{\Phi_1}}\right) \sinh{(\bar{\gamma}_{\Phi_1}\bar{L})}  \\
 \cosh{(\bar{\gamma}_{\varphi_1}\bar{L})}[2 \bar{k}^2 {\bar{\mu}_1}-{\bar{\alpha}_1 \bar{T}_1} {\bar{K}_1} {\bar{\mathscr{T}}_{\varphi_1}}-{\bar{k}^2_{\varphi_1}}
   \bar{\Gamma}_1] & \cosh{(\bar{\gamma}_{\vartheta_1}\bar{L})}[2 \bar{k}^2 {\bar{\mu}_1}-{\bar{\alpha}_1 \bar{T}_1} {\bar{K}_1}
   {\bar{\mathscr{T}}_{\vartheta_1}}-{\bar{k}^2_{\vartheta_1}} \bar{\Gamma}_1] & -2 \mi
   {\bar{\gamma}_{\Phi_1}} \bar{k} {\bar{\mu}_1}\cosh{(\bar{\gamma}_{\Phi_1}\bar{L})} \\
 \mi \bar{k} \cosh{(\bar{\gamma}_{\varphi_1}\bar{L})} & \mi \bar{k} \cosh{(\bar{\gamma}_{\vartheta_1}\bar{L})} & {\bar{\gamma}_{\Phi_1}} \cosh{(\bar{\gamma}_{\Phi_1}\bar{L})}\\
 {\bar{\gamma}_{\varphi_1}} \sinh{(\bar{\gamma}_{\varphi_1}\bar{L})} & {\bar{\gamma}_{\vartheta_1}}  \sinh{(\bar{\gamma}_{\vartheta_1}\bar{L})}& -\mi \bar{k} \sinh{(\bar{\gamma}_{\Phi_1}\bar{L})}\\
 {\bar{\mathscr{T}}_{\varphi_1}} \cosh{(\bar{\gamma}_{\varphi_1}\bar{L})} & {\bar{\mathscr{T}}_{\vartheta_1}} \cosh{(\bar{\gamma}_{\vartheta_1}\bar{L})} & 0
   \\
 {\bar{\gamma}_{\varphi_1}} {\bar{\mathscr{K}}_1} {\bar{\mathscr{T}}_{\varphi_1}} \sinh{(\bar{\gamma}_{\varphi_1}\bar{L})} & {\bar{\gamma}_{\vartheta_1}}
   {\bar{\mathscr{K}}_1} {\bar{\mathscr{T}}_{\vartheta_1}} \sinh{(\bar{\gamma}_{\vartheta_1}\bar{L})}& 0 \\
\end{array} 
\right.\\ & \qquad\qquad \left.
\begin{array}{ccc} \label{TVE SYM MATRIX A_S}
 2 \mi {\bar{\gamma}_{ \varphi_2}} \bar{k} {\bar{\mu}_2} & 2 \mi {\bar{\gamma}_{\vartheta_2}} \bar{k} {\bar{\mu}_2} & -{\bar{\mu}_2} \left(2
   \bar{k}^2 - {\bar{k}^2_{\Phi_2}}\right) \\
   -2 \bar{k}^2 {\bar{\mu}_2}+{\bar{\alpha}_2 \bar{T}_2} {\bar{K}_2}
   {\bar{\mathscr{T}}_{\varphi_2}}+{\bar{k}^2_{\varphi_2}} \bar{\Gamma}_2 & -2 \bar{k}^2 {\bar{\mu}_2}+{\bar{\alpha}_2 \bar{T}_2} {\bar{K}_2} {\bar{\mathscr{T}}_{\vartheta_2}}+{\bar{k}^2_{\vartheta_2}} \bar{\Gamma}_2 & -2 \mi {\bar{\gamma}_{\Phi_2}} \bar{k} {\bar{\mu}_2} \\
   -\mi \bar{k} & -\mi \bar{k} & {\bar{\gamma}_{\Phi_2}} \\
    {\bar{\gamma}_{\varphi_2}} &
   {\bar{\gamma}_{\vartheta_2}} & \mi \bar{k} \\
    -{\bar{\mathscr{T}}_{\varphi_2}} & -{\bar{\mathscr{T}}_{\vartheta_2}} & 0 \\  {\bar{\gamma}_{\varphi_2}} {\bar{\mathscr{K}}_2}
   {\bar{\mathscr{T}}_{\varphi_2}} & {\bar{\gamma}_{\vartheta_2}} {\bar{\mathscr{K}}_2} {\bar{\mathscr{T}}_{\vartheta_2}} & 0 \\
\end{array} \right) = \bar{\bm{A}}_{Sym}. \numberthis
 \end{align*}
 When implementing this numerically however, we find that we must factorise (\ref{TVE SYM MATRIX A_S}) columnwise to achieve bounded trigonometric functions which is acceptable since it does not change the determinant. As a result, the DE that is passed to the solver is
\begin{align*}  
    \det & \left(
\begin{array}{ccc}
 2 \mi {\bar{\gamma}_{\varphi_1}} \bar{k} {\bar{\mu}_1} \tanh{(\bar{\gamma}_{\varphi_1}\bar{L})} & 2 \mi {\bar{\gamma}_{\vartheta_1}} \bar{k} {\bar{\mu}_1} \tanh{(\bar{\gamma}_{\vartheta_1}\bar{L})}&
   {\bar{\mu}_1} \left(2 \bar{k}^2-{\bar{k}^2_{\Phi_1}}\right) \tanh{(\bar{\gamma}_{\Phi_1}\bar{L})}  \\
 2 \bar{k}^2 {\bar{\mu}_1}-{\bar{\alpha}_1 \bar{T}_1} {\bar{K}_1} {\bar{\mathscr{T}}_{\varphi_1}}-{\bar{k}^2_{\varphi_1}}
   \bar{\Gamma}_1 & 2 \bar{k}^2 {\bar{\mu}_1}-{\bar{\alpha}_1 \bar{T}_1} {\bar{K}_1}
   {\bar{\mathscr{T}}_{\vartheta_1}}-{\bar{k}^2_{\vartheta_1}} \bar{\Gamma}_1 & -2 \mi
   {\bar{\gamma}_{\Phi_1}} \bar{k} {\bar{\mu}_1} \\
 \mi \bar{k} & \mi \bar{k} & {\bar{\gamma}_{\Phi_1}} \\
 {\bar{\gamma}_{\varphi_1}} \tanh{(\bar{\gamma}_{\varphi_1}\bar{L})} & {\bar{\gamma}_{\vartheta_1}}  \tanh{(\bar{\gamma}_{\vartheta_1}\bar{L})}& -\mi \bar{k} \tanh{(\bar{\gamma}_{\Phi_1}\bar{L})}\\
 {\bar{\mathscr{T}}_{\varphi_1}}& {\bar{\mathscr{T}}_{\vartheta_1}} & 0
   \\
 {\bar{\gamma}_{\varphi_1}} {\bar{\mathscr{K}}_1} {\bar{\mathscr{T}}_{\varphi_1}} \tanh{(\bar{\gamma}_{\varphi_1}\bar{L})} & {\bar{\gamma}_{\vartheta_1}}
   {\bar{\mathscr{K}}_1} {\bar{\mathscr{T}}_{\vartheta_1}} \tanh{(\bar{\gamma}_{\vartheta_1}\bar{L})}& 0 \\
\end{array} 
\right.\\ & \qquad\qquad \left.
\begin{array}{ccc} \label{TVE-TVE-SYM}
 2 \mi {\bar{\gamma}_{ \varphi_2}} \bar{k} {\bar{\mu}_2} & 2 \mi {\bar{\gamma}_{\vartheta_2}} \bar{k} {\bar{\mu}_2} & -{\bar{\mu}_2} \left(2
   \bar{k}^2 - {\bar{k}^2_{\Phi_2}}\right) \\
   -2 \bar{k}^2 {\bar{\mu}_2}+{\bar{\alpha}_2 \bar{T}_2} {\bar{K}_2}
   {\bar{\mathscr{T}}_{\varphi_2}}+{\bar{k}^2_{\varphi_2}} \bar{\Gamma}_2 & -2 \bar{k}^2 {\bar{\mu}_2}+{\bar{\alpha}_2 \bar{T}_2} {\bar{K}_2} {\bar{\mathscr{T}}_{\vartheta_2}}+{\bar{k}^2_{\vartheta_2}} \bar{\Gamma}_2 & -2 \mi {\bar{\gamma}_{\Phi_2}} \bar{k} {\bar{\mu}_2} \\
   -\mi \bar{k} & -\mi \bar{k} & {\bar{\gamma}_{\Phi_2}} \\
    {\bar{\gamma}_{\varphi_2}} &
   {\bar{\gamma}_{\vartheta_2}} & \mi \bar{k} \\
    -{\bar{\mathscr{T}}_{\varphi_2}} & -{\bar{\mathscr{T}}_{\vartheta_2}} & 0 \\  {\bar{\gamma}_{\varphi_2}} {\bar{\mathscr{K}}_2}
   {\bar{\mathscr{T}}_{\varphi_2}} & {\bar{\gamma}_{\vartheta_2}} {\bar{\mathscr{K}}_2} {\bar{\mathscr{T}}_{\vartheta_2}} & 0 \\
\end{array} \right) = 0. \numberthis
 \end{align*}


Similarly, for the antisymmetric modes we seek solutions of the form

\begin{subequations} \label{TVE ANTISymm. Waveguides}
\begin{align} \label{TVE ANTISymm. Waveguide Compressional}
\bar{\varphi}_1 &= \bar{\text{P}1}_\text{A} \sinh{(\bar{\gamma}_{\varphi_1} \bar{y})} \me^{\mi \bar{k} \bar{x}},  &
 \bar{\varphi}_2(\bar{x},\bar{y}) = 
\begin{cases}
		  -\bar{\text{P}2}_\text{A}\me^{  \bar{\gamma}_{\varphi_2} (\bar{y}+\bar{L}) + \mi \bar{k} \bar{x}}, \qquad & \bar{y} \leq -\bar{L}\\
          \bar{\text{P}2}_\text{A}\me^{  -\bar{\gamma}_{\varphi_2} (\bar{y}-\bar{L}) + \mi \bar{k} \bar{x}}, \qquad & \bar{y} \geq  \bar{L}
\end{cases},\\ \label{TVE ANTISymm. Waveguide THERMAL}
\bar{\vartheta}_1 &= \bar{\text{T}1}_\text{A} \sinh{(\bar{\gamma}_{\vartheta_1} \bar{y})} \me^{\mi \bar{k} \bar{x}},  &
 \bar{\vartheta}_2(\bar{x},\bar{y}) = 
\begin{cases}
		  -\bar{\text{T}2}_\text{A}\me^{  \bar{\gamma}_{\vartheta_2} (\bar{y}+\bar{L}) + \mi \bar{k} \bar{x}}, \qquad & \bar{y} \leq -\bar{L}\\
          \bar{\text{T}2}_\text{A}\me^{  -\bar{\gamma}_{\vartheta_2} (\bar{y}-\bar{L}) + \mi \bar{k} \bar{x}}, \qquad & \bar{y} \geq  \bar{L}
\end{cases},\\ \label{TVE ANTISymm. Waveguide Shear}
\bar{\Phi}_1 &= \bar{\text{S}1}_\text{A} \cosh{(\bar{\gamma}_{\Phi_1} \bar{y})} \me^{\mi \bar{k} \bar{x}},  &
\bar{\Phi}_2(\bar{x},\bar{y}) = 
\begin{cases}
		  \bar{\text{S}2}_\text{A}\me^{  \bar{\gamma}_{\Phi_2} (\bar{y}+\bar{L}) + \mi \bar{k} \bar{x}}, \qquad & \bar{y} \leq -\bar{L}\\
          \bar{\text{S}2}_\text{A}\me^{  -\bar{\gamma}_{\Phi_2} (\bar{y}-\bar{L}) + \mi \bar{k} \bar{x}}, \qquad & \bar{y} \geq  \bar{L}
\end{cases},
\end{align}
\end{subequations}
for some complex valued amplitudes $\bar{\text{P}1}_\text{A}, \bar{\text{P}2}_\text{A},\bar{\text{T}1}_\text{A}, \bar{\text{T}2}_\text{A}, \bar{\text{S}1}_\text{A},  \bar{\text{S}2}_\text{A}$. The resulting matrix  $\bm{\bar{A}}_{ASym}$ is then identical to (\ref{TVE SYM MATRIX A_S}) but with $\cosh{(\cdot)}$ replaced by $\sinh{(\cdot)}$ and viceversa and is therefore not included for brevity. The dispersion equation implemented in the numerics is given by

\begin{align*} 
   & \det  \left(
\begin{array}{ccc}
 2 \mi {\bar{\gamma}_{\varphi_1}} \bar{k} {\bar{\mu}_1}  & 2 \mi {\bar{\gamma}_{\vartheta_1}} \bar{k} {\bar{\mu}_1} &
   {\bar{\mu}_1} \left(2 \bar{k}^2-{\bar{k}^2_{\Phi_1}}\right)\\
 \tanh{(\bar{\gamma}_{\varphi_1}\bar{L})} [2 \bar{k}^2 {\bar{\mu}_1}-{\bar{\alpha}_1 \bar{T}_1} {\bar{K}_1} {\bar{\mathscr{T}}_{\varphi_1}}-{\bar{k}^2_{\varphi_1}}
   \bar{\Gamma}_1 ] & \tanh{(\bar{\gamma}_{\vartheta_1}\bar{L})}[2 \bar{k}^2 {\bar{\mu}_1}-{\bar{\alpha}_1 \bar{T}_1} {\bar{K}_1}
   {\bar{\mathscr{T}}_{\vartheta_1}}-{\bar{k}^2_{\vartheta_1}} \bar{\Gamma}_1] & -2 \mi
   {\bar{\gamma}_{\Phi_1}} \bar{k} {\bar{\mu}_1} \tanh{(\bar{\gamma}_{\Phi_1}\bar{L})}\\
 \mi \bar{k} \tanh{(\bar{\gamma}_{\varphi_1}\bar{L})} & \mi \bar{k} \tanh{(\bar{\gamma}_{\vartheta_1}\bar{L})} & {\bar{\gamma}_{\Phi_1}} \tanh{(\bar{\gamma}_{\Phi_1}\bar{L})} \\
 {\bar{\gamma}_{\varphi_1}} & {\bar{\gamma}_{\vartheta_1}}  & -\mi \bar{k} \\
 {\bar{\mathscr{T}}_{\varphi_1}} \tanh{(\bar{\gamma}_{\varphi_1}\bar{L})}& {\bar{\mathscr{T}}_{\vartheta_1}} & 0
   \\
 {\bar{\gamma}_{\varphi_1}} {\bar{\mathscr{K}}_1} {\bar{\mathscr{T}}_{\varphi_1}} & {\bar{\gamma}_{\vartheta_1}}
   {\bar{\mathscr{K}}_1} {\bar{\mathscr{T}}_{\vartheta_1}} & 0 \\
\end{array} 
\right.\\ & \qquad\qquad \left.
\begin{array}{ccc} \label{TVE-TVE-ANTISYM}
 2 \mi {\bar{\gamma}_{ \varphi_2}} \bar{k} {\bar{\mu}_2} & 2 \mi {\bar{\gamma}_{\vartheta_2}} \bar{k} {\bar{\mu}_2} & -{\bar{\mu}_2} \left(2
   \bar{k}^2 - {\bar{k}^2_{\Phi_2}}\right) \\
   -2 \bar{k}^2 {\bar{\mu}_2}+{\bar{\alpha}_2 \bar{T}_2} {\bar{K}_2}
   {\bar{\mathscr{T}}_{\varphi_2}}+{\bar{k}^2_{\varphi_2}} \bar{\Gamma}_2 & -2 \bar{k}^2 {\bar{\mu}_2}+{\bar{\alpha}_2 \bar{T}_2} {\bar{K}_2} {\bar{\mathscr{T}}_{\vartheta_2}}+{\bar{k}^2_{\vartheta_2}} \bar{\Gamma}_2 & -2 \mi {\bar{\gamma}_{\Phi_2}} \bar{k} {\bar{\mu}_2} \\
   -\mi \bar{k} & -\mi \bar{k} & {\bar{\gamma}_{\Phi_2}} \\
    {\bar{\gamma}_{\varphi_2}} &
   {\bar{\gamma}_{\vartheta_2}} & \mi \bar{k} \\
    -{\bar{\mathscr{T}}_{\varphi_2}} & -{\bar{\mathscr{T}}_{\vartheta_2}} & 0 \\  {\bar{\gamma}_{\varphi_2}} {\bar{\mathscr{K}}_2}
   {\bar{\mathscr{T}}_{\varphi_2}} & {\bar{\gamma}_{\vartheta_2}} {\bar{\mathscr{K}}_2} {\bar{\mathscr{T}}_{\vartheta_2}} & 0 \\
\end{array} \right) = 0. \numberthis
 \end{align*}
It is easy to check that for $\bar{k}\bar{L} \gg 1$ (\ref{TVE-TVE-SYM}), (\ref{TVE-TVE-ANTISYM}) reduce to (\ref{stoneley:TVE-TVE}). 
\subsection{A note on direct comparisons between TVE and VE solutions}
Care must be taken when comparing solutions between TVE and VE theories, especially with \textit{fluid} media (as is the case in e.g.\ Figures \ref{Fig:LeakyRayleigh}c), \ref{Fig:Stoneley/Scholte}c)). This is the case since, in the absence of thermal effects, the standard acoustic/elastic models are based on an adiabatic consideration as opposed to isothermal \cite{nowacki2013thermoelasticity}, as is the case for TVE, and therefore the corresponding moduli must be adjusted accordingly if they are to be compared directly. By manipulating thermodynamic relations, it is easy to show that
\begin{equation}\label{Adiab vs Iso Moduli}
    K_\text{Adiab} = \frac{c_p}{c_v} K_\text{Iso}=\gamma K_\text{Iso}, \qquad  \mu_\text{Adiab} = \mu_\text{Iso}
\end{equation}
see e.g.\ \cite{lubarda2004thermodynamic} (where also equivalent expression for $E, \nu, \lambda$ are derived). Since $\gamma$ is close to $1$ for many common solids and therefore from (\ref{Adiab vs Iso Moduli}) the resulting material parameters remain close, this distinction is often not made in elasticity. Nevertheless, in studies of the type presented here with the unified TVE model, it is paramount to make the distinction for the validity of the results, see Table \ref{table: Th parameters}.

\end{appendices}
\end{document}